\documentclass[prd,aps,twocolumn,nofootinbib, superscriptaddress]{revtex4-1}

\usepackage{color}
\usepackage{times}

\usepackage[utf8]{inputenc}
\usepackage{amsmath}
\usepackage{amssymb}
\usepackage{graphicx}
\usepackage{bm}

\usepackage{ulem}
\usepackage{hyperref}
\usepackage{thumbpdf} 
\usepackage{enumerate}
\usepackage{graphicx}
\usepackage{graphics,epsfig,placeins,wrapfig}
\usepackage{braket}
\usepackage{longtable}
\usepackage{hhline}

\def\be{\begin{equation}}
\def\en{\end{equation}}
\def\bea{\begin{eqnarray}}
\def\ena{\end{eqnarray}}

\def\msun{\,{\rm M_\odot}}

\newcommand{\vS}{\mbox{\boldmath${S}$}}
\newcommand{\vR}{\mbox{\boldmath${r}$}}
\newcommand{\vp}{\mbox{\boldmath${p}$}}
\newcommand{\vL}{\mbox{\boldmath${L}$}}
\newcommand{\vN}{\mbox{\boldmath${N}$}}
\newcommand{\vLhat}{\mbox{\boldmath${\hat{L}}$}}

\newcommand{\vJ}{\mbox{\boldmath${J}$}}
\newcommand{\vE}{\mbox{\boldmath${\hat{e}}$}}

\begin{document}

\title{Validating the effective-one-body model of spinning, precessing binary black holes against numerical relativity}

\author{Stanislav Babak}
\affiliation{Max Planck Institute for Gravitational Physics (Albert Einstein Institute), Am M\"uhlenberg 1, Potsdam-Golm, 14476, Germany}
\author{Andrea Taracchini}
\affiliation{Max Planck Institute for Gravitational Physics (Albert Einstein Institute), Am M\"uhlenberg 1, Potsdam-Golm, 14476, Germany}
\author{Alessandra Buonanno}
\affiliation{Max Planck Institute for Gravitational Physics (Albert Einstein Institute), Am M\"uhlenberg 1, Potsdam-Golm, 14476, Germany}
\affiliation{Department of Physics, University of Maryland, College Park, MD 20742, USA}

\date{\today}

\begin{abstract}
 In Ref.~\cite{Abbott:2016izl}, the properties of the first 
  gravitational wave detected by LIGO, GW150914, were measured by employing an
  effective-one-body (EOB) model of precessing binary black holes whose 
underlying dynamics and waveforms were calibrated to numerical-relativity (NR) simulations. 
Here, we perform the first extensive
  comparison of such EOBNR model to 70 precessing NR waveforms that span mass ratios 
from 1 to 5, dimensionless spin magnitudes up to 0.5, generic spin orientations, and length of
  about 20 orbits. We work in the observer's inertial frame and 
  include all $\ell=2$ modes in the gravitational-wave
  polarizations. We introduce new prescriptions for the EOB ringdown
  signal concerning its spectrum and time of onset. For total masses
  between $10\msun$ and $200\msun$, we find that precessing EOBNR waveforms 
 have  unfaithfulness within about 3\% to NR waveforms 
when considering the Advanced-LIGO design noise curve. This result is
  obtained without recalibration of the inspiral-plunge of the
  underlying  nonprecessing EOBNR model. The unfaithfulness is computed with maximization over time and phase of
  arrival, sky location and polarization of the EOBNR waveform and it is averaged over
  sky location and polarization of the NR signal. We
  also present comparisons between NR and EOBNR waveforms in a frame that tracks the orbital
  precession.
\end{abstract}

\pacs{04.30.-w, 04.25.-g, 04.25.D-, 04.25.dg}

\maketitle

\section{Introduction}
Starting from 2011 the Laser Interferometer Gravitational-wave Observatory (LIGO)~\cite{TheLIGOScientific:2014jea} underwent a major upgrade to its advanced configuration, and from September 2015 to January 2016 it operated at an unprecedented sensitivity, nearly $\sim 3$ times better than that of the initial configuration. During this period, LIGO confidently detected two stellar-mass binary black hole (BBH) coalescences, GW150914~\cite{Abbott:2016blz} and GW151226~\cite{Abbott:2016nmj}, thus inaugurating the age of GW astronomy. These discoveries provide the evidence for the existence of BBHs~\cite{TheLIGOScientific:2016src}, and hint at the possibility that they could dominate the event rate~\cite{TheLIGOScientific:2016pea}.

BBH searches and parameter-estimation analyses are based on matched filtering, which relies on accurate theoretical models of the expected GW signal emitted by these systems. In order not to bias the estimation of the source's physical parameters, it is crucial to develop waveform models that can describe the most generic binary configuration, namely one where the spins of the two BHs do not have any preferential alignment with the orbital angular momentum, and thus induce a precession of the orbital plane with respect to a fixed inertial frame. There exist two main astrophysical channels that lead to the formation of stellar-mass BBHs~\cite{lrr-2014-3, lrr-2013-4}: (i) from isolated stellar binaries, as the end product of stellar (co)evolution, and (ii) from the dynamical capture in dense stellar environments. The masses and spins of a BBH depend on the specific formation channel, so that their measurement via GW observations could shed light on the formation mechanism~\cite{Vitale:2015tea}. 

In this paper we adopt the effective-one-body (EOB) approach~\cite{Buonanno:1998gg, Buonanno:2000ef} to model the orbital dynamics and the inspiral-merger-ringdown GW emission of coalescing compact-object binaries in the time domain. This model combines results from different realms of relativity (post-Newtonian theory, BH perturbation theory, gravitational self force, numerical relativity) into a unified framework with the goal of bridging the gaps present between the respective domains of validity of those methods. Over the last decade, a lot of effort has been put into (i) including new analytical information provided by the most recent theoretical developments in the general-relativistic 2-body problem, and (ii) calibrating the model to numerical-relativity (NR) simulations to improve its accuracy. State-of-the-art EOB models for spinning, nonprecessing BBHs are described in Refs.~\cite{Taracchini:2013rva,Nagar:2015xqa}. 

For the first time, Ref.~\cite{Pan:2013rra} proposed a complete EOBNR model for spinning, precessing BBHs. The idea is to use a precessing frame that tracks the motion of the orbital plane~\cite{Buonanno:2002fy}. In this frame, precession-induced amplitude and phase modulations in the waveforms are minimized, thus one employs a nonprecessing EOBNR model calibrated to NR to generate inspiral-plunge modes. The modes are then rotated to the inertial frame that is aligned with the spin of the remnant BH: in this frame, the merger-ringdown modes are generated and smoothly stitched to the rotated inspiral-plunge modes. Without recalibration of the underlying nonprecessing EOBNR model~\cite{Taracchini:2012ig}, the model was compared to two long, precessing NR simulations, finding very good agreement. 

In this paper we consider the precessing EOBNR model of Ref.~\cite{Pan:2013rra} (henceforth referred to as ``precessing EOBNR"), that uses the EOBNR model of Ref.~\cite{Taracchini:2013rva} as the underlying nonprecessing model. We introduce changes that concern the modeling of the merger-ringdown signal (i.e., time of onset and spectrum). This model is completely generic: it depends on \textit{all} six spin degrees of freedom, and can generate waveforms for any mass ratio, arbitrary dimensionless spin magnitudes, and arbitrary spin orientations. This model was used to infer the properties of GW150914 in Ref.~\cite{Abbott:2016izl}, finding consistent results with Ref.~\cite{TheLIGOScientific:2016wfe}, which employed a nonprecessing EOBNR model~\cite{Taracchini:2013rva} and a precessing frequency-domain inspiral-merger-ringdown phenomenological model~\cite{Hannam:2013oca} (henceforth referred to as ``precessing IMRPhenom''). Here, we carry out the first large-scale comparison of precessing EOBNR waveforms to 70 public NR waveforms of precessing BBHs produced by the Simulating eXtreme Spacetime (SXS) collaboration~\cite{Mroue:2013xna}. Using the zero-detuned high-power Advanced-LIGO design noise curve~\cite{Shoemaker2009, Aasi:2013wya}, we find remarkable agreement for total masses in the range $10\msun$ -- $200\msun$, with a sky- and polarization-averaged unfaithfulness $\lesssim 3\%$ for inclinations $0,\pi/3$.  We find that the 
performance slightly degrades as we approach edge-on configurations (a few configurations reach a maximum unfaithfulness between 3\% and 4\%), but in all cases the level of agreement is suitable for detection of such binaries in Advanced LIGO at its design sensitivity. In the context of parameter estimation, it is reasonable to expect that an unfaithfulness on the order of $3\%$ will not induce systematic errors that are larger than the statistical errors, whose size is set by the finite signal-to-noise ratio of the signals. We also check one of the main assumptions in the precessing EOBNR model, namely the possibility of approximating the precessing-frame GW modes with nonprecessing GW modes. Notably, we employ the maximum-radiation frame of Refs.~\cite{Schmidt:2010it, O'Shaughnessy:2011fx, Boyle:2011gg, Schmidt:2012rh} as the waveform-based precessing frame, and verify that in that frame the NR and EOBNR waveforms agree very well during the inspiral-plunge phase. 

This paper is organized as follows. In Sec.~\ref{S:NPmodel}, we review the EOBNR model for nonprecessing BBHs. In Sec.~\ref{S:model} we describe the precessing EOBNR model and in Sec.~\ref{S:chip} we point out the main differences with respect to the precessing 
IMRPhenom model. In Sec.~\ref{S:comparison} we discuss the comparison of the precessing EOBNR model to 70 precessing NR waveforms.  In Sec.~\ref{S:p-frame} we assess the impact of some of the approximations used in the model by studying waveforms in the precessing frame. Section~\ref{S:concl} summarizes our main conclusions. In Appendix~\ref{S:rot} we provide formulas to rotate GW modes from 
the inertial to the precessing frame. In Appendix~\ref{AppendixA} we describe the procedure to compute the \textit{averaged}  
unfaithfulness between two generic, precessing waveforms.  In Appendices~\ref{AppendixB} and~\ref{AppendixC} we 
summarize new prescriptions for the EOB ringdown signal that improve the stability of the model with respect to 
Ref.~\cite{Pan:2013rra}.

In what follows, we use geometric units $G=1=c$. Given any 3-vector $\boldsymbol{u}$, we indicate its norm with $u \equiv \sqrt{\boldsymbol{u}\cdot \boldsymbol{u}}$, and we define the respective unit vector as $\boldsymbol{\hat{u}}\equiv \boldsymbol{u}/u$.

\section{Effective-one-body model of nonprecessing binary black holes}
\label{S:NPmodel}

This section heavily relies on Refs.~\cite{Taracchini:2012ig, Taracchini:2013rva}, wherein more details can be found.

A BBH coalescence spans a large range of regimes, from the slow-motion, weak-field inspiral stage, to the highly dynamical, strong-field merger phase, to the relaxation to a single, stationary Kerr BH. The EOB formalism can account for the GW emission of the entire process by combining both analytical and numerical results that describe those different stages of the coalescence. Let $m_{1,2}$ be the two BH masses (with $m_1 \geq m_2$) and $\vS_{1,2} \equiv m_{1,2}^2 \boldsymbol{\chi}_{1,2}$ their spins. The EOB approach relies on a Hamiltonian $H_{\textrm{EOB}}$ that encodes the BBH conservative dynamics~\cite{Buonanno:1998gg, Buonanno:2000ef}. Two are the main building blocks: (i) the Hamiltonian $H_{\textrm{eff}}$ of a spinning particle of mass $\mu \equiv m_1 m_2/(m_1 + m_2)$ and spin $\vS_* \equiv \vS_*(m_1,m_2,\vS_1,\vS_2)$ moving in an effective, deformed Kerr spacetime of mass $M\equiv m_1 + m_2$ and spin $\vS_{\textrm{Kerr}} \equiv \vS_1 + \vS_2$~\cite{Barausse:2009aa,Barausse:2009xi,Barausse:2011ys}; (ii) an energy mapping between $H_{\textrm{eff}}$ and $H_{\textrm{EOB}}$~\cite{Buonanno:1998gg}
\be
H_{\textrm{EOB}} \equiv M\sqrt{1+2\nu\left(\frac{H_{\textrm{eff}}}{\mu} - 1\right)}-M\,, 
\en 
where $\nu \equiv \mu/M$ is the symmetric mass ratio. The deformation of the effective Kerr metric is fixed by requiring that $H_{\textrm{EOB}}$ agrees with the PN Hamiltonian for BBHs in the low frequency regime. It is found that the deformation away from Kerr is regulated by $\nu$ alone. The spin-orbit (spin-spin) couplings are incorporated in the model through 3.5PN (2PN) order~\cite{Barausse:2009xi,Barausse:2011ys}. The dynamical variables in the EOB model are the relative separation $\vR$ (pointing from body 2 to body 1) and its canonically conjugate momentum $\vp$, and the spins $\vS_{1,2}$. The conservative EOB dynamics is completely general and can naturally accommodate precession.

If, at some reference time, the BH spins are exactly aligned or antialigned with the Newtonian orbital angular momentum $\vL_N\equiv \mu\, \vR \times \dot{\vR}$, then the orbital plane is fixed with respect to an inertial observer, the spins are constant vectors, and we speak of a \textit{nonprecessing} BBH. On the other hand, when the BH spins have generic orientations, both the orbital plane and the spins undergo precession about the total angular momentum of the binary, defined as $\vJ \equiv \vL + \vS_1 + \vS_2$, where $\vL \equiv \mu\, \vR \times \vp$; in this case, we speak of a \textit{precessing BBH}.

Here we focus on the nonprecessing case. The emission of GWs causes a BBH to lose angular momentum proportionally to the GW flux. The EOB model accounts for this dissipation through a nonconservative radiation-reaction force~\cite{Buonanno:2005xu,Damour:2007xr,Damour:2008gu,Pan:2010hz}
\be
\boldsymbol{\mathcal{F}} \equiv \frac{\Omega}{16\pi} \frac{\vp}{|\vL|}\sum_{\ell=2}^{8}\sum_{m=-\ell}^{\ell} m^2\vert D_{\textrm{L}} h_{\ell m}\vert^2\,,
\en
where $\Omega \equiv |\vR \times \dot{\vR}|/|\vR|^2$ is the orbital frequency, $D_{\textrm L}$ is the luminosity distance of the BBH to the observer, and the $h_{\ell m}$'s are the GW modes.\footnote{As usual, the GW polarizations $h_{+,\times}$ are combined into the combination $h_{+} - i h_{\times}$, which is then projected onto $-2$-spin-weighted spherical harmonics, thus obtaining the individual multipolar modes $h_{\ell m}$.} Although ready-to-use, frequency-domain PN formulas for the GW modes exist in the literature~\cite{Arun:2008kb,Mishra:2016whh}, the EOB model employs a factorized, resummed version of them~\cite{Damour:2007xr,Damour:2008gu,Pan:2010hz}. The resummation was developed only for quasicircular, nonprecessing BBHs, and improves the accuracy of the PN expressions in the test-particle limit, as demonstrated by comparisons to numerical solutions of the frequency-domain Regge-Wheeler-Zerilli and Teukolsky equations. 

Thus, the inspiral-plunge GW emission is modeled in two steps: first, one numerically integrates Hamilton's equations for $H_{\textrm {EOB}}(\vR,\vp,\vS_1,\vS_2)$ subject to the dissipative force $\boldsymbol{\mathcal{F}}$ from quasispherical initial conditions~\cite{Buonanno:2005xu} down to the light-ring (or photon orbit) crossing; second, one evaluates the factorized, resummed waveforms $h_{\ell m}$ on the orbital dynamics obtained in the previous step.

The description of a BBH as a system composed of two individual objects is of course valid only up to the merger. After that point, the EOB model builds the GW emission (ringdown stage) via a superposition of quasinormal modes (QNMs) of the remnant BH that forms after the coalescence of the progenitors~\cite{Buonanno:2000ef}. The transition from the inspiral-plunge to the ringdown stage roughly occurs around the time when the light ring of $H_{\textrm{EOB}}$ is crossed, consistently with the picture of QNMs leaking outside the gravitational potential barrier of the newly formed BH~\cite{Davis:1971gg,Davis:1972ud}. QNM frequencies and decay times are known (tabulated) functions of the mass $M_f$ and spin $\vS_f \equiv M_f^2 \boldsymbol{\chi}_f$ of the remnant BH~\cite{Berti:2005ys}. In the original papers~\cite{Buonanno:2000ef,Buonanno:2005xu,Buonanno:2006ui}, both $M_f$ and $\vS_f$ were derived internally to the model, without any input from NR. The ringdown signal is smoothly joined to the inspiral-plunge signal (see Ref.~\cite{Taracchini:2012ig}), thus completing the construction of the full waveform.

The EOB prediction of the inspiral-merger-ringdown waveform predated the first NR computations of 2005 (that considered equal-mass, nonspinning BBHs)~\cite{Pretorius:2005gq,Campanelli:2005dd,Baker:2005vv}, and was found to be in good qualitative agreement with them~\cite{Buonanno:2006ui}. The analysis of ground-based GW interferometric data puts rather stringent requirements on how accurately templates should reproduce fully general relativistic waveforms~\cite{Lindblom:2010mh}. Over the past decade, a lot of effort has been put into improving the performance of the EOB model against NR. In particular, in recent years the model has incorporated inputs coming from: (i) new higher-order PN computations of the conservative dynamics up to 4PN order (for nonspinning terms)~\cite{Damour:2015isa},  (ii) new higher-order PN computations of the GW modes up to 2PN spin-spin terms~\cite{Buonanno:2012rv}, (iii) gravitational-self-force computations of conservative effects like the innermost-stable circular-orbit shift~\cite{Barack:2009ey}, (iv) BH perturbation theory computations in Kerr spacetime of GW fluxes and merger waveforms~\cite{Barausse:2011kb,Taracchini:2013wfa,Taracchini:2014zpa,Harms:2014dqa,Harms:2015ixa}, and (v) direct tuning of unknown high-order PN parameters, nonquasicircular corrections, and remnant properties to NR simulations of nonprecessing BBHs~\cite{Taracchini:2012ig,Taracchini:2013rva,Nagar:2015xqa}.

Building on Ref.~\cite{Pan:2013rra}, in Sec.~\ref{S:model} we will include spin-precession effects in the nonprecessing EOB model of Ref.~\cite{Taracchini:2013rva}. The latter was calibrated to 38 NR simulations of coalescing BBHs with mass ratios between 1 and 8, dimensionless BH spin magnitudes up to 0.98 (0.5) for equal-mass (unequal-mass) systems, and about 10 to 30 orbits long. Only the dominant $(2,2)$ mode was tuned to NR. This model is known as \verb+SEOBNRv2+ in the LIGO Algorithm Library (LAL). In the low-frequency regime, Refs.~\cite{Szilagyi:2015rwa,Kumar:2015tha} successfully tested the model with new nonspinning and nonprecessing-spins NR simulations that exceeded the typical length of the NR waveforms employed for its calibration. Reference~\cite{Kumar:2016dhh} compared the model to 95 new NR waveforms~\cite{Chu:2015kft} of aligned-spin BBHs of typical length, finding performances that are quite good for mild spin magnitudes ($\lesssim 0.5$) and worse for spin-magnitudes about 0.8 and mass ratios 2-3, where the model is extrapolated. It is important to bear in mind that in this paper we consider precessing BBHs with only mild spin magnitudes, so that any large-spin-magnitude inaccuracy in the underlying nonprecessing model is not a concern. Work is underway to retune the nonprecessing EOBNR model to the new set of NR waveforms of Ref.~\cite{Chu:2015kft}, while trying to improve its extrapolation properties~\cite{SEOBNRv4}.

\section{Effective-one-body model of precessing binary black holes}
\label{S:model}
Here we review and improve the precessing EOBNR model. The discussion heavily relies 
on Ref.~\cite{Pan:2013rra}, wherein more details can be found. 

\subsection{Inspiral-plunge waveforms}
As explained in Sec.~\ref{S:NPmodel}, the conservative dynamics of the EOB model can handle generic, spin-precessing BBHs. In particular, BH spin precession is described by the following equation of motion
\be
\frac{\mathrm{d}\vS_{1,2}}{\mathrm{d}t} = \frac{\partial H_{\mathrm{EOB}}}{\partial \vS_{1,2}} \times \vS_{1,2}\,.
\en

Some care must be taken when modeling the dissipation of energy and angular momentum into GWs for the generation of the inspiral-plunge dynamics. The radiation-reaction force $\boldsymbol{\mathcal{F}}$ that supplements the equations of motion built from the conservative Hamiltonian depends on the amplitude of the individual GW modes $|h_{\ell m}|$, whose factorized, resummed frequency-domain expressions have been derived only for nonprecessing BBHs, and are functions -- among other parameters -- of the (constant, in this case) aligned-spin magnitudes $\boldsymbol{\chi}_{1,2}\cdot \vLhat_N$. In the precessing case, the natural choice is therefore to allow for time dependence, letting the modes depend on $\boldsymbol{\chi}_{1,2}(t)\cdot \vLhat_N(t)$. Moreover, we let the GW modes depend on the generic, precessing EOB orbital dynamics through the radial separation $r$ and orbital frequency $\Omega$, which carry spin-spin modulations whenever precession is present. Of course, not all spin-precession imprints are accounted for, as is clearly seen by inspection of the PN formulas in Refs.~\cite{Arun:2008kb,Mishra:2016whh}, where an additional, explicit dependence on the in-plane spin components is shown. 
\begin{figure}
\includegraphics[angle=0,width=\linewidth]{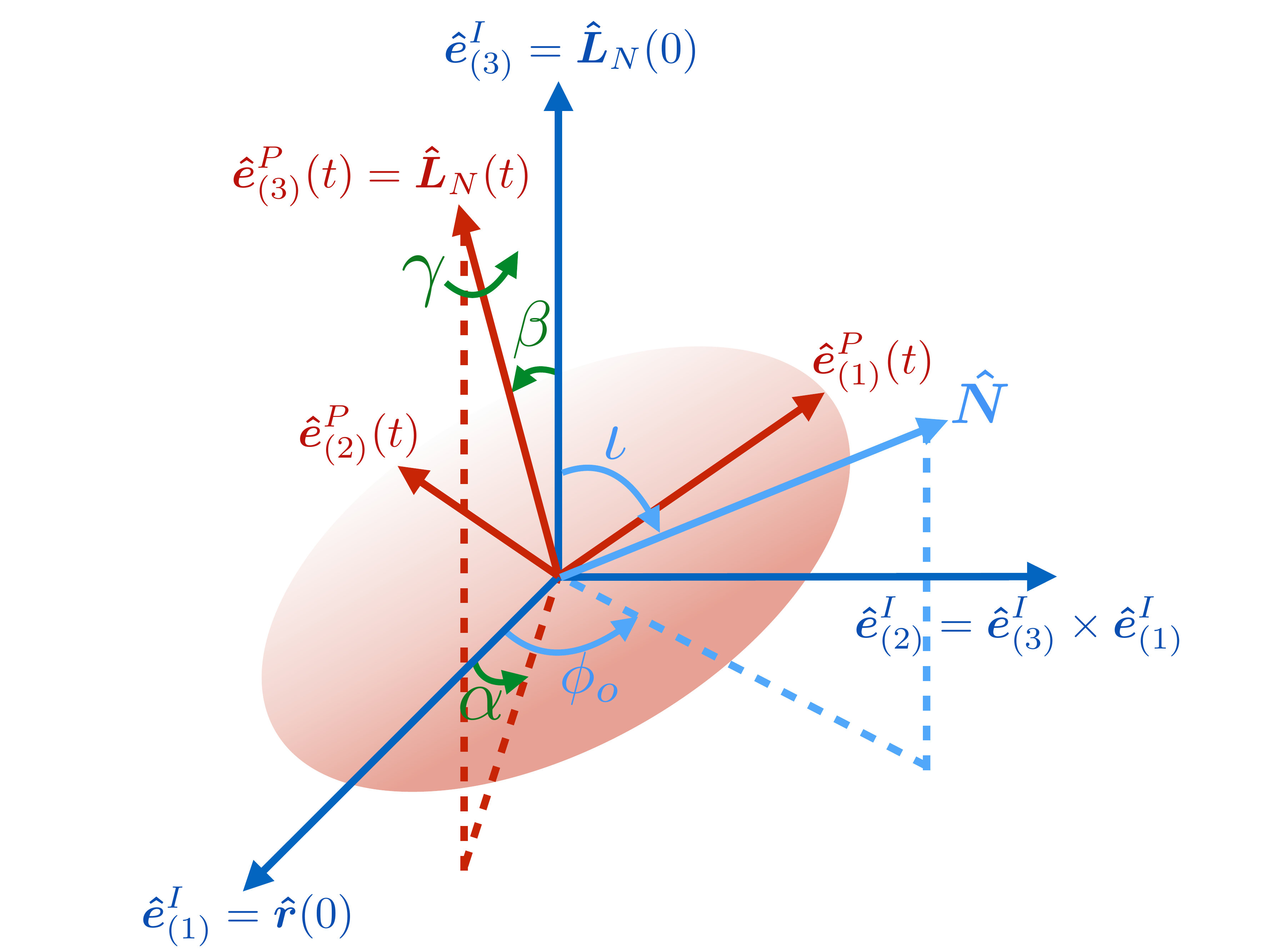}
\caption{\textbf{Observer's frame}, defined by the directions of the initial Newtonian angular momentum $\vL_N(0)$ and separation $\vR(0)$, and \textbf{precessing frame}, instantaneously aligned with $\vL_N(t)$ and described by the Euler angles $(\alpha, \beta, \gamma)$ (see Eq.~(\ref{EulerAngles})).}\label{F:Pframe} 
\end{figure}  

For the purpose of data analysis, one is typically interested in computing the GW polarizations as seen by an inertial-frame observer. We call this frame the \textit{observer's frame} and denote quantities expressed in such frame with the superscript $I$. In particular, the observer's frame is described by the triad $\{\vE^I_{(i)}\}$ ($i=1,2,3$), where $\vE_{(1)}^I\equiv\boldsymbol{\hat{r}}(0)$, $\vE^I_{(3)} \equiv \vLhat_N(0)$ and $\vE_{(2)}^I\equiv\vE_{(3)}^I \times \vE_{(1)}^I$. In this frame, the line of sight of the observer is parametrized as $\boldsymbol{\hat{N}} \equiv (\sin{\iota}\cos{\phi_o},\sin{\iota}\sin{\phi_o},\cos{\iota})$ (see Fig.~\ref{F:Pframe}). The observer's frame should be complemented by a polarization basis, that spans the plane orthogonal to $\boldsymbol{\hat{\vN}}$. We choose polarization basis vectors $\{\vE_{(1)}^r,\vE_{(2)}^r\}$ such that $\vE_{(1)}^r \equiv (\vE^I_{(3)}\times\boldsymbol{\hat{N}})/|\vE^I_{(3)}\times\boldsymbol{\hat{N}}|$ and $\vE_{(2)}^r \equiv \boldsymbol{\hat{N}} \times \vE_{(1)}^r$.

We now outline how the observer's-frame modes $h_{\ell m}^I$ are computed in the precessing EOBNR model. Let us first discuss the inspiral-plunge, for which a precessing EOB dynamics is available. During this portion of the BBH coalescence, one can define a non-inertial reference frame that tracks the motion of the orbital plane. We refer to this frame as \textit{precessing frame} (superscript $P$), and describe it with the triad $\{\vE^P_{(i)}\}$ ($i=1,2,3$). At each instant, its $z$-axis is aligned with $\vLhat_N$: $\vE_{(3)}^P \equiv \vLhat_N(t)$. In this frame, the BBH is viewed face-on at all times, and the GW radiation is channelled mostly into the $(2,\pm 2)$ modes, which look very much alike nonprecessing waveforms~\cite{Buonanno:2002fy,Schmidt:2010it,Boyle:2011gg,O'Shaughnessy:2011fx,Schmidt:2012rh}. The other two axes lie in the orbital plane and are defined as to minimize precessional effects in the precessing-frame modes $h_{\ell m}^P$~\cite{Buonanno:2002fy,Boyle:2011gg}. After introducing the vector $\boldsymbol{\Omega}_e \equiv \vLhat_N\times {\textrm d}\vLhat_N/{\textrm d}t$, the minimum-rotation condition is enforced by ${\textrm d}\vE_{(1),(2)}^P/{\textrm d}t = \boldsymbol{\Omega}_e\times \vE_{(1),(2)}^P$ and $\vE_{(1),(2)}^P(0) = \vE^I_{(1),(2)}$ (see also Fig.~\ref{F:Pframe}). We parametrize the rotation from the precessing to the observer's frame by means of time-dependent Euler angles $(\alpha(t),\beta(t),\gamma(t))$ that are computed using Eqs.~(\ref{EulerAngles})--(\ref{EulerAnglesSpecial2}) in Appendix~\ref{S:rot}. Note that the minimum-rotation condition can also be expressed as a differential equation for $\gamma$: $\dot{\gamma} = -\dot{\alpha}\cos{\beta}$ with $\gamma(0)=-\alpha(0) = \pi/2$. 

\begin{figure*}
\includegraphics[angle=0,width=\linewidth, keepaspectratio=true]{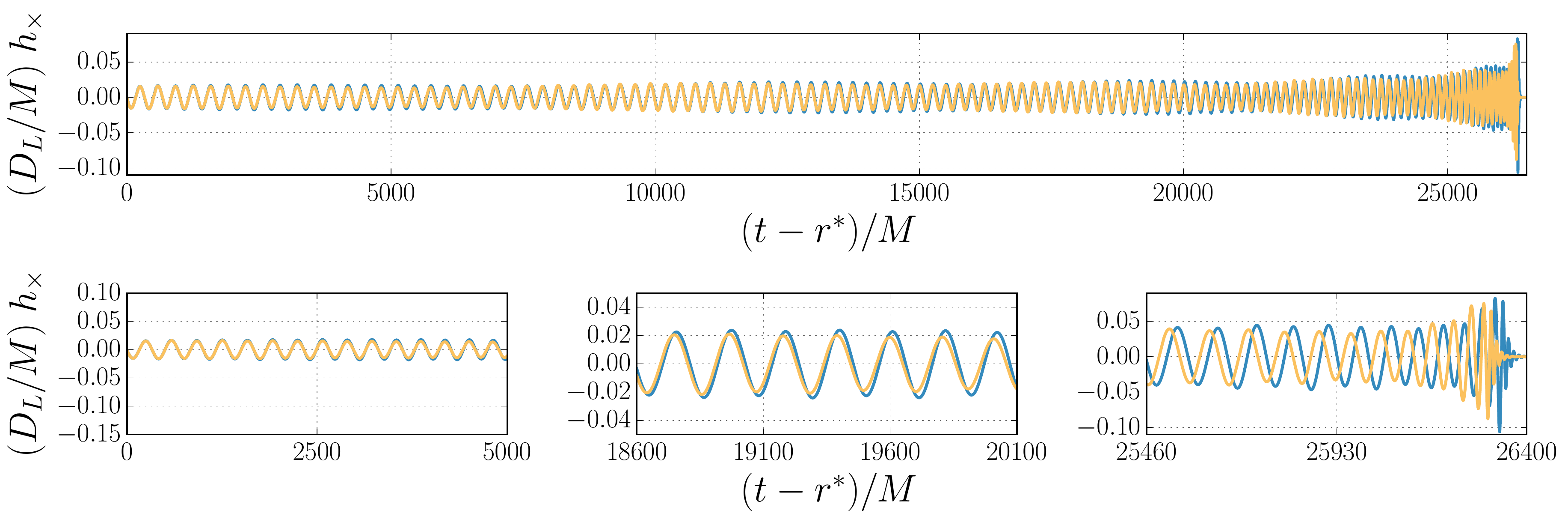} 
 \caption{\textbf{Impact of the six independent spin degrees of freedom on the waveform}: we compare two precessing EOBNR waveforms that differ only by the opening angle between the initial in-plane spins. The initial intrinsic parameters of the two configurations are $q=3$, $\boldsymbol{\chi}_1\cdot\vLhat_N =\boldsymbol{\chi}_2\cdot\vLhat_N = 0$, $\chi_{1\perp} = 0.01$, $\chi_{2\perp} = 0.9$, and different opening angles between the in-plane spins, $104^{\circ}$ and $171^{\circ}$, respectively.}
 \label{F:chip}
\end{figure*}

Following the idea that in the precessing frame the GW emission resembles that of a nonprecessing BBH, we model the precessing-frame inspiral-plunge modes just like we do for the GW flux, namely by using the factorized, resummed nonprecessing waveforms and evaluating them along the EOB precessing dynamics up to the light-ring crossing time and employing the time-dependent spin projections $\boldsymbol{\chi}_{1,2}(t)\cdot \vLhat_N(t)$. Next, the observer's-frame inspiral-plunge modes are obtained by rotating the precessing-frame inspiral-plunge modes with Eq.~(\ref{hlmRotation}). We include only $\ell = 2$ modes with $h^P_{20} = 0$ and $h^P_{2-m} = h^{P*}_{2m}$. This approximation was investigated in NR studies~\cite{Pekowsky:2013ska,Boyle:2014ioa} that found the asymmetries between opposite-$m$ modes to be small as compared to the dominant $(2,2)$-mode emission (at least during the inspiral) in a corotating frame that maximizes emission in the $(2,\pm 2)$ modes, also known as \textit{maximum-radiation frame}~\cite{Boyle:2011gg,Boyle:2013nka}. However, the difference in phase and amplitude between positive and negative $m$-modes might become non-negligible at merger. By construction, the $z$-axis of the maximum-radiation frame of the precessing EOBNR model coincides with the $z$-axis of the precessing frame.

As noted in Ref.~\cite{Pan:2013rra},  the time evolution of $\vL$ is much simpler than that of $\vL_N$: while the EOB angular momentum $\vL$ simply precesses about the total angular momentum $\boldsymbol{J}$, the Newtonian angular momentum $\vL_N$ also displays angular nutations at twice the orbital frequency, a behavior that is accounted for by simple PN considerations~\cite{Kidder:1995zr, Schmidt:2010it}.

\subsection{Ringdown waveforms}
Just like for nonprecessing BBHs, the modeling of the ringdown signal for precessing BBHs is done with a linear combination of QNMs of the remnant BH. We remind that the ringdown construction consists of (i) a prescription for the final mass and spin of the remnant BH,  (ii) a choice of which QNMs should be superposed, and (iii) a matching procedure to the inspiral-plunge signal. Let $t_{\textrm{match}}$ be the time when the ringdown signal begins and is stitched to the inspiral-plunge waveform. As mentioned earlier, $t_{\textrm{match}}$ is close to the light-ring crossing; for the precise, improved prescriptions adopted here, see Appendices~\ref{AppendixB} and \ref{AppendixC}. 

For the mass and spin magnitude of the remnant we adopt the same phenomenological fits to NR that are part of the nonprecessing model and that are functions of $(\nu,\boldsymbol{\chi}_1\cdot\vLhat, \boldsymbol{\chi}_2\cdot\vLhat)$; for precessing systems, we simply evaluate the spin projections at $t=t_{\textrm{match}}$. Note that we choose projections onto $\vLhat$ instead of $\vLhat_N$ to avoid the nutations present in the latter. One important difference with respect to the nonprecessing case is that now we also have to model the direction of the final spin, not just its magnitude: we assume that it is parallel to the total angular momentum at the ringdown onset $\boldsymbol{\hat{\chi}}_f = \boldsymbol{\hat{J}}(t_{\textrm{match}})$. 

As to the QNM spectrum, when modeling the ringdown of the $(\ell,m)$ spherical mode in nonprecessing EOB models the typical choice is to employ the first 8 overtones of the QNM labelled by spheroidal indices $(\ell,m)$~\cite{Berti:2005ys}. 
When $\boldsymbol{\hat{L}}(t_{\textrm{match}})$ is almost aligned with $\boldsymbol{\hat{\chi}}_f$ the spectrum of the $(2,m)$ ringdown signal is dominated by $(2, |m|,n)$ QNMs of the remnant, while in the case of strong antialignment $(2, -|m|,n)$ QNMs are the most excited.  The precise frequency content of the ringdown is discussed in Appendices~\ref{AppendixB} and \ref{AppendixC}. 
 The ringdown signal reads 
\be
\sum_{n=0}^8 A_n \exp{[(i\, \omega_{\ell m n}-1/\tau_{\ell m n})(t-t_{\textrm{match}})]}\,, 
\en
where $\omega_{\ell mn}>0$ and $\tau_{\ell mn}>0$ are the (real) frequency and decay time of the $(\ell,m,n)$ overtone, and the $A_n$'s are fixed by the matching procedure~\cite{Taracchini:2012ig} to the inspiral-plunge signal. Since the QNMs are defined with respect to the direction of the final spin, the specific form of the ringdown signal, as a linear combination of QNMs, is formally valid only in an inertial frame whose $z$-axis is parallel to $\boldsymbol{\hat{\chi}}_f$. Thus, we perform the stitching of the ringdown to the inspiral-plunge signal in a frame, that we call \textit{attachment frame} and denote with a $J$ superscript, such that $\vE^J_{(3)} \equiv \boldsymbol{\hat{\chi}}_f$; the other axes are $\vE^J_{(1)} \equiv( \vE_{(3)}^J \times \vE_{(3)}^I)/| \vE_{(3)}^J \times \vE_{(3)}^I|$ and  $\vE_{(2)}^J \equiv \vE_{(3)}^J \times \vE_{(1)}^J$. If 
$\vE^J_{(3)}$ is parallel to $\vE_{(3)}^I$ (spin are (anti)aligned),  then $\vE^J_{(1)} \equiv (\vE_{(3)}^J\cdot\vE_{(3)}^I)  \vE_{(1)}^I$. We rotate the inspiral-plunge GW modes $h^P_{2 m}$ from the precessing frame, where they are generated, to the attachment frame using the rotation formula in Eq.~(\ref{hlmRotation}) and the Euler angles that parametrize the rotation from one frame to the other as in Eqs.~(\ref{EulerAngles})--(\ref{EulerAnglesSpecial2}), and obtain the inspiral-plunge modes $h_{2 m}^J$. We attach the ringdown waveforms to the inspiral-plunge $h^J_{2 m}$'s, and get complete inspiral-merger-ringdown waveforms. Finally, we compute the inspiral-merger-ringdown modes $h^I_{2m}$ in the observer's frame by rotating the inspiral-merger-ringdown $h^J_{2 m}$'s. The observer's-frame polarizations then read
\be
h_+(t, \iota,\phi_o) - ih_{\times}(t,\iota,\phi_o) = \sum_{ m=-2}^{2} h^I_{2 m}(t)\, _{-2}Y_{2 m}(\iota,\phi_o)\,,
\en
where we denote with $\iota$ the inclination angle and with $\phi_o$ the azimuthal direction to the observer.

\section{Differences with the phenomenological model of precessing binary black holes}
\label{S:chip}

References~\cite{Schmidt:2014iyl,Hannam:2013oca} proposed a precessing inspiral-merger-ringdown frequency-domain model\footnote{In the LIGO Algorithm Library this approximant is known as \texttt{IMRPhenomPv2}.} (precessing IMRPhenom) based on the nonprecessing phenomenological model of Ref.~\cite{Khan:2015jqa}. Similarly to what is done for the precessing EOBNR model, the inertial-frame waveforms are generated by rotating the nonprecessing modes according to the precessional motion of the orbital plane. The Euler angles parametrizing the rotation that connects the precessing frame to the observer's frame are derived from PN theory.  The underlying nonprecessing modes depend on the BH masses and the projections of the two BH spins onto the orbital angular momentum, with the in-plane spin components entering only in the formula for the spin of the remnant BH. Precessional effects are regulated by the initial phase of the binary in the orbital plane and a single spin parameter,

\be
\chi_p \equiv \frac{1}{B_1 m_1^2} \max{(B_1 S_{1 \perp}, B_2 S_{2 \perp})}\,,\label{chip}
\en
where $B_1 \equiv 2 + 3m_2/(2m_1)$ and $B_2 \equiv 2 + 3m_1/(2m_2)$ (with $m_1 \geq m_2$), and the $S_{i\perp}\equiv m_i^2 \chi_{i\perp}$ ($i=1,2$) are the components of the spins perpendicular to the orbital angular momentum. Therefore in the precessing IMRPhenom model the six spin degrees of freedom are approximated by only four independent parameters~\cite{Schmidt:2014iyl}. On the other hand, the precessing EOBNR model depends on all six spin components. In Fig.~\ref{F:chip} we compare two precessing EOBNR waveforms that have the same mass ratio ($q = 3$), the same initial spin projections on $\vLhat_N$ ($\boldsymbol{\chi}_1\cdot\vLhat_N =\boldsymbol{\chi}_2\cdot\vLhat_N = 0$), the same initial spin projections on the orbital plane ($\chi_{1\perp} = 0.01$, $\chi_{2\perp} = 0.9$), but different opening angle between the in-plane spins ($104^{\circ}$ and $171^{\circ}$, respectively). We see from Eq.~(\ref{chip}) that these two waveforms share the same $\chi_p$, however they are quite different, both in phase and amplitude modulations, especially close to merger. The precessing IMRPhenom model predicts identical waveforms for these configurations. We should point out that an unfaithfulness comparison (see Sec.~\ref{S:comparison}) of the two waveforms gives values around 7\% (0.4\%) without (with) sky-location and polarization averaging. This result implies that, from the point of view of a parameter-estimation study, it is conceivable that the difference observed in Fig.~\ref{F:chip} can be absorbed by a bias in the polarization angle and in the azimuthal position of the observer (with the appropriate shift in time). Those parameters are typically poorly measurable from a GW observation with two LIGO detectors.

There exist other salient differences between the precessing EOBNR and IMRPhenom models: (i) The inspiral of the precessing-frame modes is obtained from the purely nonprecessing model in the IMRPhenom model, while the precessing EOBNR model employs the fully precessing dynamics to generate the $h_{2m}^P$'s; (ii) The ringdown signal is generated in the frame of the remnant BH for the precessing EOBNR model -- where the QNMs from BH perturbation theory are computed -- while the IMRPhenom model builds it in the precessing frame; (iii) The IMRPhenom model contains only $m=\pm 2$ modes in the precessing frame, while the precessing EOBNR model includes (uncalibrated) $m=\pm 1$ modes; (iv) The IMRPhenom model employs a stationary-phase approximation to the mode-rotation formula, while the precessing EOBNR model performs the rotations in the time domain according to Eq.~(\ref{hlmRotation}); (v) While the EOBNR Euler angles are calculated from the generic motion of the EOBNR orbital angular momentum, the IMRPhenom Euler angles are obtained from approximations concerning the spin and frequency evolution and from the high-frequency extension of PN results.

Depending on the region of parameter space, the differences between the two precessing models may or may not be relevant. In particular, in the case of GW150914 -- an almost equal-mass, face-off, non-extremal BBH -- Ref.~\cite{Abbott:2016izl} showed that both approximants give consistent estimations of the parameters of the source. 

Lastly, it is worth mentioning that the approximated description of the precessional spin dynamics in IMRPhenom -- as well as the frequency-domain formulation -- entails a significant computational speedup of the phenomenological model as compared to precessing EOBNR. Ongoing work~\cite{v3ROM} is applying reduced-order-modeling techniques to precessing EOBNR with the goal of speeding up waveform generation by several orders of magnitude, as done in the past for spin-aligned EOBNR models~\cite{Purrer:2014fza,Purrer:2015tud}.

\section{Comparisons to numerical relativity in the inertial frame of an observer}
\label{S:comparison}

We now compare precessing EOBNR and NR waveforms in the observer's inertial frame.

\subsection{Numerical-relativity simulations}

\begin{figure*}
\includegraphics[angle=0,width=0.8\linewidth]{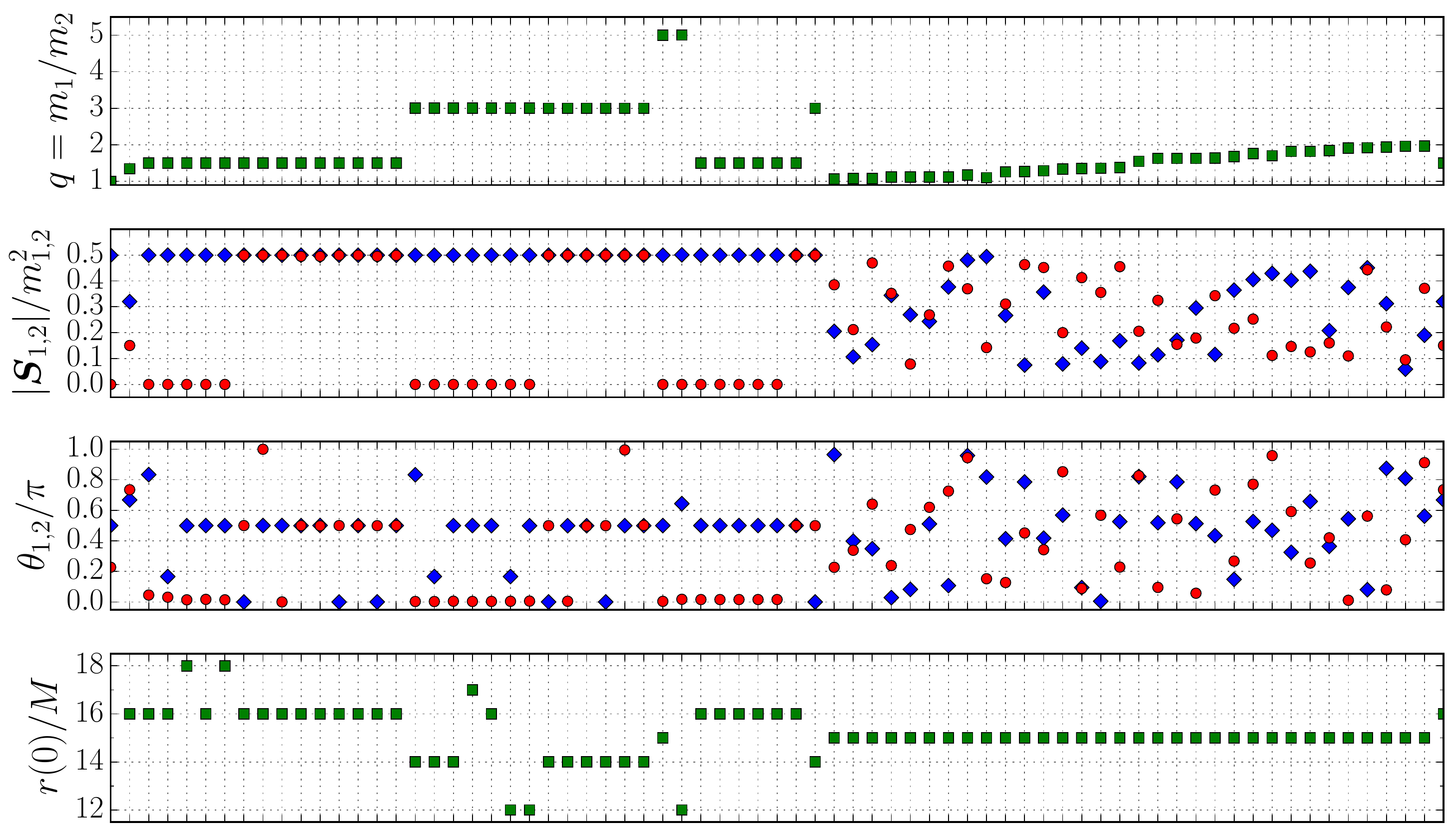}
\caption{\textbf{Parameters of the SXS NR waveforms} that are compared to the precessing EOBNR model. The horizontal axis indicates the SXS ID number of the simulation. In the four panels we show mass ratio $q$, dimensionless spin magnitudes $|\boldsymbol{\chi}_{1,2}|$, spin opening angles $\theta_{1,2} \equiv \arccos{(\boldsymbol{\hat{\chi}}_{1,2} \cdot \vLhat_N)}$, and orbital separation $r$ measured at relaxation time from the NR data. Blue points are for BH 1 (the heavier), while red ones are for BH 2.}
\label{F:SXS_prec_runs}
\end{figure*}

We consider NR waveforms from the public catalog of the SXS collaboration~\cite{Mroue:2013xna}. We 
use the NR dynamics data to read out the spins and orbital configuration at a
time after the junk radiation has left the system (i.e., after the
relaxation time, as reported in the SXS catalog). We select 70
precessing BBH simulations with initial coordinate separation larger
than $12$ M and eccentricity $\le 10^{-3}$. We summarize the parameters of these simulations in
Fig.~\ref{F:SXS_prec_runs} where the horizontal axis is just indexing
the
simulations.
We show mass ratio, dimensionless spin magnitudes, spin opening angles
with respect to the Newtonian angular momentum, and coordinate
separation measured at the relaxation time from the NR
data. Information about the heavier (lighter) BH is in blue diamond
(red circle). Most simulations have moderate mass ratios, with only
two at mass ratio 5. Only moderate dimensionless spin magnitudes are
covered (up to 0.5), and 20 out of 70 runs have spin only on the
heavier object. Initial spin opening angles are random for almost half
of the simulations, while many of the single-spin runs initially have
$\vS_1$ in the orbital plane. Most runs span around 20 orbits. For
each run, we choose the highest resolution that is available and $N=4$
extrapolation order. For comparisons to the precessing
EOBNR model, we use only the $\ell=2$ modes.

\subsection{Waveform alignment and unfaithfulness for precessing binaries}
Let us consider an L-shaped GW detector and define the \textit{detector frame} as the one whose $z$-axis is orthogonal to the plane of the detector, while the $x$- and $y$-axes are lined up with its arms. A GW emitted by the coalescence of a precessing BBH will cause a strain in the detector that depends on 15 independent parameters: the BH masses $m_{1,2}$, the initial BH spin vectors $\vS_{1,2}$, the angular position of the line of sight in the source's frame $(\iota, \phi_o)$, the sky location of the source in the detector frame $(\theta, \phi)$, the polarization angle $\psi$, the luminosity distance of the source $D_{\textrm{L}}$, the time of arrival $t_c$.\footnote{In the literature, the time of arrival has several definitions, and is usually associated with the time at which the GW signal reaches a certain frequency or its maximum amplitude.} We consider a parameter to be \textit{extrinsic} if it defines the position and the orientation of the source frame with respect to the detector frame, while an \textit{intrinsic} parameter is defined only in the source frame. Note that the intrinsic parameters also define the waveform subspace where one places templates for template-based GW searches. 
For the purpose of generating a waveform, we choose the value of the GW frequency at some initial time $t_0$ and integrate the dynamics forward in time. However, for the purposes of detection and parameter estimation, it is more convenient to refer all quantities to the time of arrival $t_c$. We note that $t_0$ and $t_c$ are degenerate parameters.

The standard approach to comparing waveforms consists in computing their \textit{overlap} integral in the frequency domain with a frequency-dependent weight given by the single-sided noise power spectral density of the GW detector of interest: $(\hat{h}|\hat{s})$, where $\hat{h}$ and $\hat{s}$ are the normalized signal and template waveforms (see Appendix~\ref{AppendixA}). In our case, we treat the NR waveforms as \textit{signals} and the EOBNR waveforms as \textit{templates}. 
In search pipelines, one aims at maximizing the signal-to-noise ratio over a bank of templates that are placed as densely as possible in the BBH parameter space~\cite{Allen:2005fk,Babak:2012zx}. Thus, the common approach is to maximize the overlap over both intrinsic and extrinsic parameters of the template -- \textit{effectualness}~\cite{Damour:1997ub}. Currently, Advanced LIGO searches employ only aligned-spin template banks~\cite{TheLIGOScientific:2016qqj}, but there is a recent proposal on how to extend them to precessing-spin templates~\cite{Harry:2016ijz}. However, when comparing the precessing EOBNR model to NR, we want to adopt a more restrictive approach by keeping the intrinsic parameters fixed to the same value in both the signal and the template, while meaningfully optimizing/averaging over the extrinsic parameters.

Let us first discuss how to choose the intrinsic parameters (prescribed at the initial time $t_0$) for the precessing EOBNR model given a NR simulation -- \textit{waveform alignment}. At the relaxation time reported in the NR catalog, we measure: (i) the NR mass ratio, (ii) the magnitude and orientation of the NR spins, and (iii) the GW frequency of the NR $(2,2)$ mode. This is done in the inertial frame whose $z$-axis is aligned with the NR $\vLhat_N$ and whose $x$-axis is aligned with the NR BH coordinate separation vector at the relaxation time.\footnote{In the SXS catalog, both the GW modes and the dynamical vectors are provided in the inertial frame whose z-axis is aligned with the NR $\vLhat_N$ and whose $x$-axis is aligned with the NR BH coordinate separation vector at a time that is different from the relaxation time.} We set the EOBNR mass ratio and initial spin vectors to the same values measured in NR. The NR $(2,2)$ frequency at the relaxation time cannot be directly used to specify the initial EOBNR $(2,2)$ frequency because the NR data may display small oscillations due to (i) persistence of the junk radiation, (ii) residual orbital eccentricity,  and (iii) spin-spin couplings~\cite{Buonanno:2010yk}. Thus, we pick an initial EOBNR $(2,2)$ frequency that, while being to within 5\% of the measured NR value, rather guarantees the same time-domain length of the waveform. In particular, we require that the peak of $\sum_{m=-2}^2 |h_{2m}|^2$ occurs at the same time in NR and EOBNR, as elapsed from the relaxation time and $t_0$, respectively. This choice for the initial EOBNR frequency has the advantage of being formulated in frame-independent way. 
 \begin{figure*}
\includegraphics[angle=0,width=\linewidth, keepaspectratio=true]{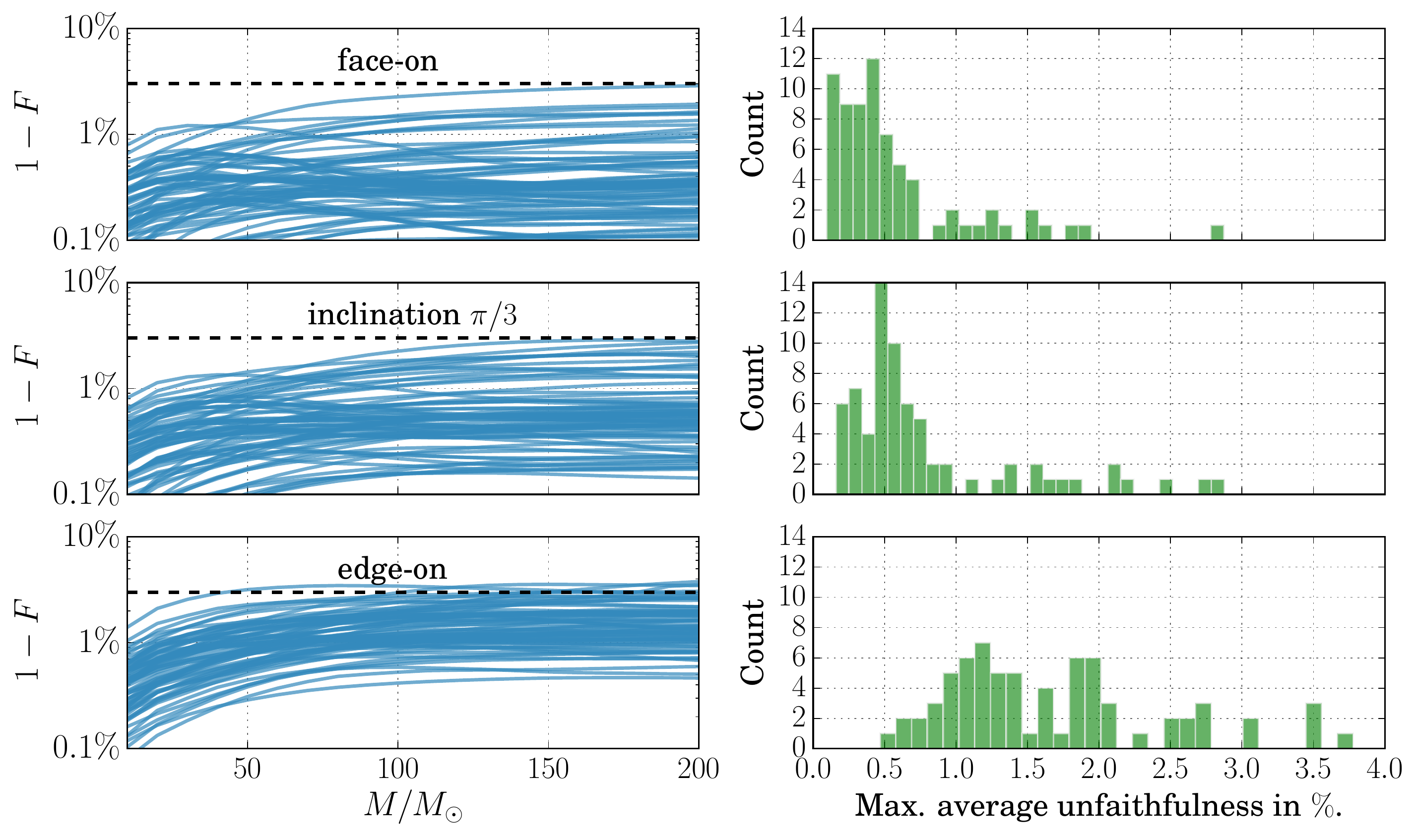} 
 \caption{\textbf{Sky- and polarization-averaged unfaithfulness} (Eq.~(\ref{EqUF})) between NR and EOBNR waveforms of 70 precessing BBHs in Advanced LIGO. \textit{Left panels}: unfaithfulness as a function of the total mass of the binary for three possible inclinations $\iota = 0,\pi/3, \pi/2$. \textit{Right panels}: histograms of the maximum unfaithfulness over the total mass range $10\msun \leq M \leq 200 \msun$.
 The dashed line corresponds to 3\% unfaithfulness.}
 \label{F:UF}
\end{figure*}

Let us now move to the extrinsic parameters. As mentioned before, the luminosity distance $D_{\textrm{L}}$ drops out of the computation. Similarly to what is done in the nonprecessing case, we numerically maximize the overlaps over $\phi_o^h$ of the template and relative difference between the $t_c$'s of template and signal. Due to the presence of $m=\pm1$ modes, the $\phi_o^h$ maximization cannot be done analytically. Both translation in time and rotation of the bodies in the orbital plane result in different binary configurations, since the spin orientations are changed with respect to the orbital separation vector. Since the evolution of the spins is rather slow when compared to the inspiral timescale, the time and phase maximizations do not significantly change the system. Furthermore, we numerically average over the $\phi_o^s$ of the NR signal. In general, we have to bear in mind that (i) a priori, the NR and EOBNR definitions of the spin vectors may be different, and (ii) $\phi_o$ and $t_c$ are not interesting from an astrophysical perspective. Concerning the inclination $\iota$, we choose it identical in template and signal. Finally, let us consider the polarization and sky-location angles. The NR and EOBNR modes are combined into the polarizations $s_{+,\times}$ and $h_{+,\times}$, respectively, according to Eq.~(\ref{E:modes}), with $\ell=2$. The polarizations are then combined into the observed strains according to
\bea
s &= F_{+}(\psi_s, \theta_s, \phi_s) s_{+} &+ F_{\times}(\psi_s, \theta_s, \phi_s) s_{\times}\,,\label{E:sstrain}\\
h &= F_{+}(\psi_h, \theta_h, \phi_h) h_{+} &+ F_{\times}(\psi_h, \theta_h, \phi_h) h_{\times}\,,\label{E:hstrain}
\ena
where $F_{+,\times}$ are the antenna pattern functions (specific to the shape of the GW detector), $\psi_{s,h}$ are the polarization angles, and $\theta_{s,h}$ and $\phi_{s,h}$ are coordinates of the sky location of the source in the inertial frame of the observer.  
We compute the \textit{min-max} and \textit{max-max} overlaps of $s$ and $h$ (see Appendix~\ref{AppendixA}) by maximizing over $(\psi_h, \theta_h, \phi_h)$ while minimizing or maximizing over $(\psi_s, \theta_s, \phi_s)$, respectively.  Min-max and max-max give the worst and best overlap, respectively, across all possible sky locations and polarizations. This motivates us to \textit{average} the overlaps over $(\psi_s, \theta_s, \phi_s)$, rather than minimizing or maximizing over them thus obtaining a quantity (the \textit{average-max} overlap) that is bound by the min-max and max-max. Interestingly, we find that the average-max overlaps are always closer to the max-max overlaps. We want to stress that here we do not intend to assess systematic biases in the measurement of the extrinsic BBH parameters over which we maximize: this will be the focus of future work. We define the \textit{sky- and polarization-averaged faithfulness} as
\begin{widetext}
\be
\label{EqUF}
F =  \max_{\phi_o^h,t_c} \max_{\psi_h, \theta_h, \phi_h } \frac{1}{16 \pi^3} \int_0^{2\pi} {\textrm d}\psi_s  \int_{-1}^1 {\textrm d}(\cos\theta_s)\int_0^{2\pi} {\textrm d}\phi_s \int_0^{2\pi} {\textrm d}\phi_o^s\;
(\hat{h} | \hat{s})\,,
\en
\end{widetext}
where $\max_{t_c}$ is a shorthand for the relative-$t_c$ maximization.
All inner products are computed with the noise curve of Advanced LIGO in the zero-detuned high-power configuration that is expected for 2020~\cite{Shoemaker2009, Aasi:2013wya}.  Finally, the \textit{sky- and polarization-averaged unfaithfulness} is defined as $1-F$.

The emission of GWs is strongest from BBHs that are face-on/off (i.e., $\iota=0,\pi$), so those are the systems that are most likely to be observed (a prime example being GW150914). On the other hand, the effect of subdominant modes is suppressed in face-on/off binaries while it is emphasized by edge-on inclinations.  Thus, we consider three possible inclinations: $\iota = 0, \pi/3, \pi/2$. For total masses in the range $10\msun \leq M \leq 200\msun$, we compute the unfaithfulness of the precessing EOBNR model against the 70 NR simulations, according to the procedure outlined above. Results of the comparison are presented in Fig.~\ref{F:UF}. Note that we do not show error bars representing the NR error (due to finite resolution or waveform extrapolation) since they are typically smaller than $10^{-3}$ (see, e.g. Fig.~10 of Ref.~\cite{Pan:2013rra} and the extensive study of Ref.~\cite{Chu:2015kft}). For inclinations $\iota=0,\pi/3$, we find that the unfaithfulness is below 1\% for the majority of waveforms, with few cases lying between 1\% and 2\%, and only two cases slightly exceeding 2\%. This means that, upon maximization over other parameters like masses and spins, the precessing EOBNR model will be effectual for these 70 BBH configurations. For edge-on binaries ($\iota=\pi/2$), we find larger values of unfaithfulness, with the bulk of configurations still below 2\%, and with the worst configuration below 4\%~(SXS:BBH:0053, with mass ratio $q=3$, and initial spins $\vS_1=(0.5,0,0)$ and $\vS_2 = (-0.5,0,0)$). We remark that the inspiral-plunge portion of the underlying nonprecessing EOBNR model \verb+SEOBNRv2+~\cite{Taracchini:2013rva} was \textit{not} recalibrated to precessing NR waveforms; the only improvements in the modeling of precessing BBHs concern the ringdown spectrum and attachment (see Appendices~\ref{AppendixB} and \ref{AppendixC}). The larger unfaithfulness for edge-on BBHs is mainly due to the fact that the precessing-frame $(2,1)$ mode -- which has large impact for such inclination -- is not calibrated to any NR data. For all inclinations, we observe that most unfaithfulness curves as functions of the total mass tend to grow towards high values of $M$: this indicates some modeling inaccuracies in the merger-ringdown portion and, again, this mostly stems from the uncalibrated merger of the precessing-frame $(2,1)$ mode. One possible, future improvement could come from attaching the ringdown in the precessing frame, where modes with different $m$'s do not interfere. This should lead to a more robust attachment procedure, similarly to what happens for the nonprecessing EOBNR model~\cite{Taracchini:2012ig}.
\begin{figure*}
\includegraphics[angle=0,width=\linewidth, keepaspectratio=true]{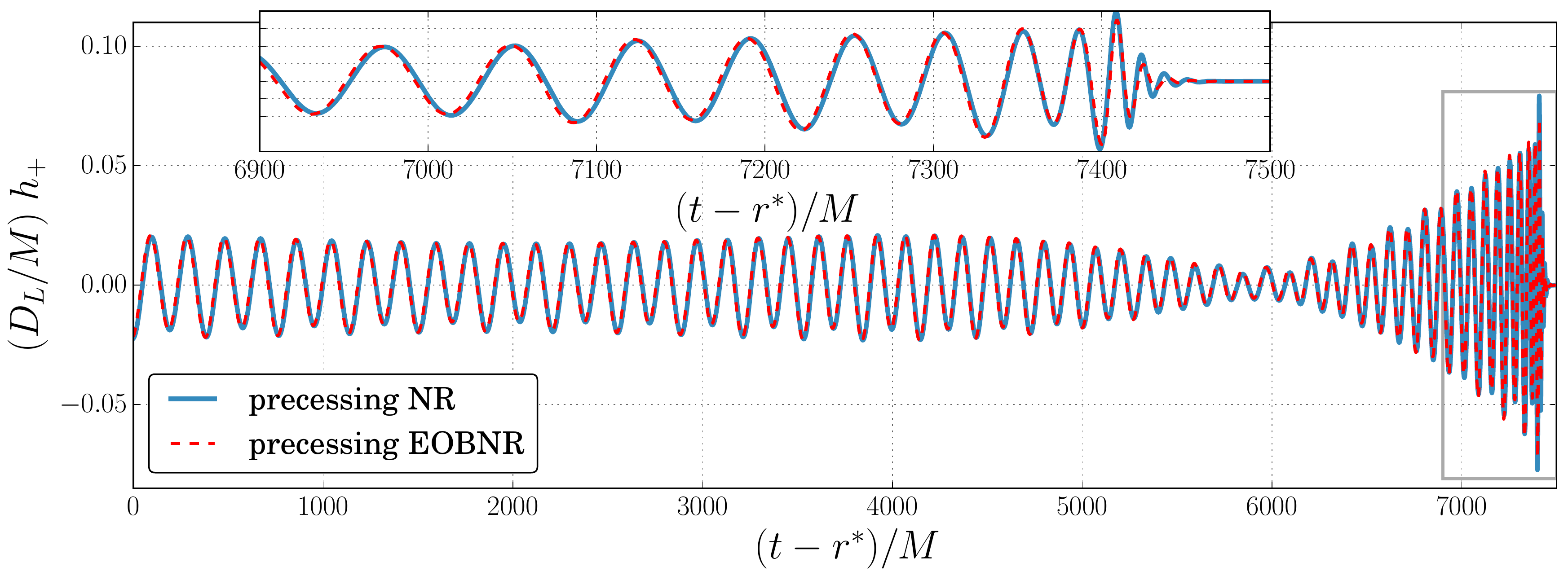}
 \caption{\textbf{Comparison of NR and EOBNR $+$ polarization} for a precessing BBH with $q=5$, $\chi_1=0.5$, $\chi_2=0$, with spin 1 initially in the orbital plane (SXS:BBH:0058). The inclination is $\iota = \pi/3$. The NR (EOBNR) data are shown in blue (dashed red).}
 \label{F:waveC58}
\end{figure*}
\begin{figure}
  \includegraphics[angle=0,width=1.\linewidth, keepaspectratio=true]{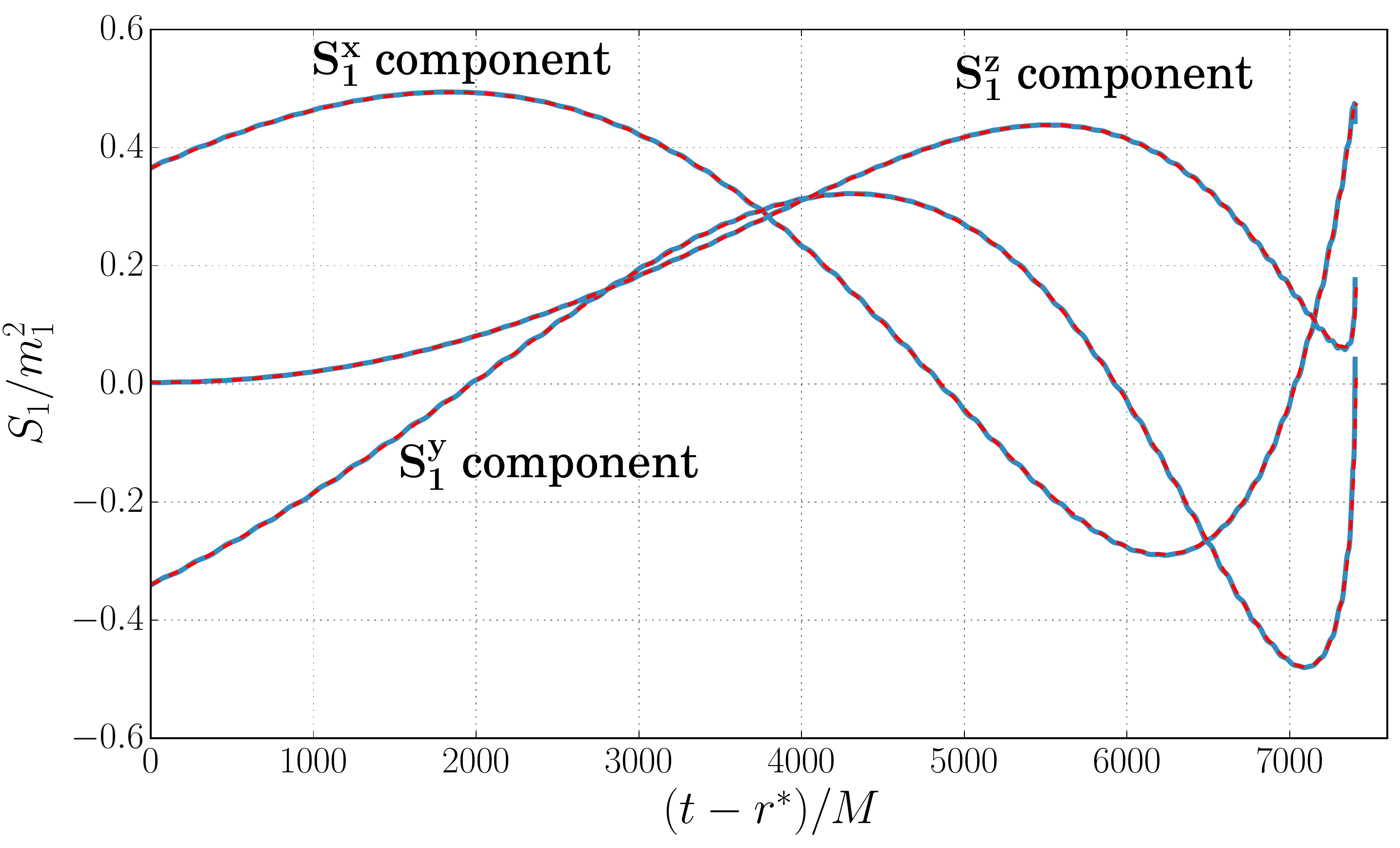}
 \caption{\textbf{Comparison of NR and EOBNR spin evolution} for the same run of Fig.~\ref{F:waveC58} (SXS:BBH:0058). We show the time evolution of the Cartesian components of $\vS_1$. The NR (EOBNR) data are shown as solid blue (dashed red) lines.}
 \label{F:spinC58}
\end{figure}

To better illustrate the level of agreement of the precessing EOBNR model to NR, we consider the SXS:BBH:0058 simulation -- a single-spin BBH with mass ratio 5 and with the heavier BH spin of dimensionless magnitude 0.5 and starting in the orbital plane of the binary. This very same run was also studied in Ref.~\cite{Pan:2013rra}. Figure~\ref{F:waveC58} shows a direct NR/EOBNR comparison of the $+$ polarizations in the time domain for $\iota = \pi/3$. The EOBNR waveform is plotted for those values of $\phi^h_o$ and $t_c$ that maximize the sky- and polarization-averaged overlap. We want to highlight the great accuracy of the model in both phasing and amplitude during the whole inspiral, consistently with what was found in Ref.~\cite{Pan:2013rra}. Also, the merger and ringdown are significantly improved with respect to Ref.~\cite{Pan:2013rra} thanks to the new prescriptions described in Appendices~\ref{AppendixB} and \ref{AppendixC}. For the same simulation, in Fig. \ref{F:spinC58} we show the evolution of the spin components in the inertial frame $\{\boldsymbol{\hat{e}}^I_{(i)}\}$, finding remarkable agreement between NR and EOBNR. Reference~\cite{Ossokine:2015vda} looked at a similar comparison between NR and PN spin dynamics.

\section{Comparison to numerical relativity in the maximum-radiation frame}
\label{S:p-frame}
The precessing EOBNR model relies on the central assumption that in the precessing frame the inspiral-plunge waveforms can be accurately modeled ignoring in-plane components of the spins. In this section we will test this assumption. This also gives us insights in how to further improve the precessing EOBNR model.  

\subsection{Maximum-radiation axis}
\label{sec:RadFrame}
\begin{figure}
  \includegraphics[angle=0,width=\linewidth, keepaspectratio=true]{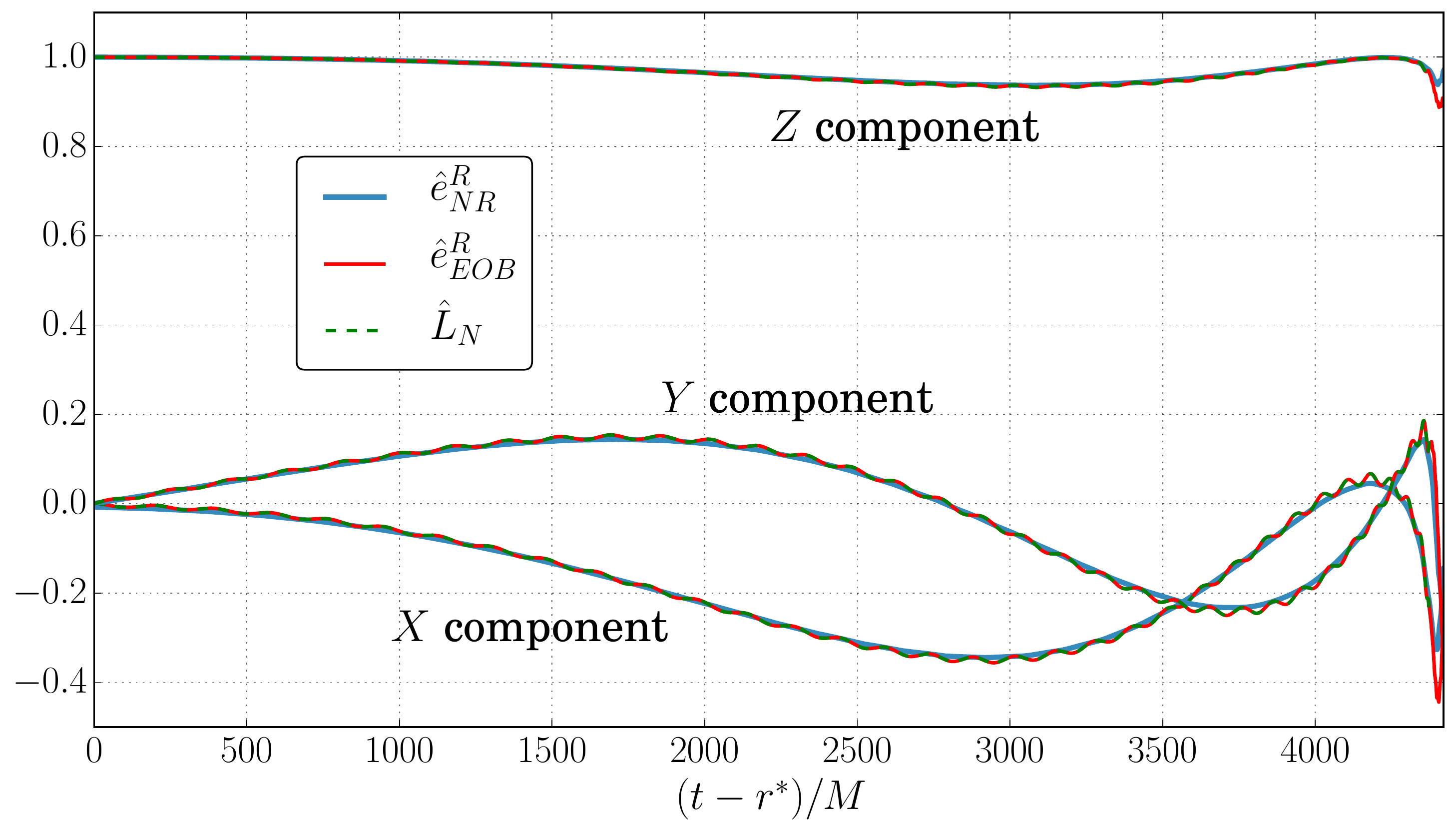}
 \caption{
 \textbf{Comparison of NR and EOBNR maximum-radiation axis} $\boldsymbol{\hat{e}}_{(3)}^R$ for a precessing BBH with $q=1.76$, $\chi_1 = 0.41$, $\chi_2= 0.25$, $\theta_1=0.53\,\pi$, and $\theta_2=0.77\,\pi$ (SXS:BBH:0137). The Cartesian components are given with respect to the inertial frame of an observer, $\{\boldsymbol{\hat{e}}^I_{(i)}\}$. The NR (EOBNR) curves are shown in blue (red). Also shown in green are the components of the EOBNR Newtonian angular momentum $\vLhat_N = \boldsymbol{\hat{e}}_{(3)}^P$.}
 \label{F:axis}
\end{figure}
As explained in Sec.~\ref{S:model}, the generation of inspiral-plunge precessing EOBNR waveforms in the inertial frame of the observer consists of two steps:
(i) generating the waveforms in the precessing frame $\{\vE^P_{(i)}\}$ using the nonprecessing formulas evaluated on the fully precessing orbital dynamics and using instantaneous projections of the spin onto $\vLhat_N$, 
and (ii) rotating the waveforms to the inertial frame according to the dynamics of $\vLhat_N$. 
 
The GW emission of a binary system is strongest along the direction orthogonal to the orbital plane. Choosing this direction~\cite{Schmidt:2010it, O'Shaughnessy:2011fx, Boyle:2011gg, Schmidt:2012rh} and adding the minimum rotation condition~\cite{Boyle:2011gg} amounts to defining a noninertial frame (\textit{maximum-radiation frame}) that relies on the radiation alone.
We follow Refs.~\cite{O'Shaughnessy:2011fx,Boyle:2011gg} and compute the maximum-radiation frame associated with EOBNR waveforms and we find that $\boldsymbol{\hat{e}}_{(3)}^P=\vLhat_N$ is identical to the direction that maximizes the strength of the GW emission, that we denote by $\boldsymbol{\hat{e}}_{(3)}^R$. 
This is not surprising since in the model $\pm m$ precessing-frame modes are symmetric and the motion of $\vLhat_N$ determines the rotation of the modes to the observer's frame. In Fig.~\ref{F:axis} we compare the radiation axis $\boldsymbol{\hat{e}}_{(3)}^R$ of EOBNR and NR waveforms for the precessing binary SXS:BBH:0137. This BBH waveform has an initial separation of about $15~M$, mass ratio 1.76, spin magnitudes $\chi_1 = 0.41$ and $\chi_2= 0.25$, and initial spin opening angles (with respect to the initial $\vLhat_N$) $\theta_1=0.53\,\pi$ and $\theta_2=0.77\,\pi$. We plot the inertial-frame components of the EOBNR maximum-radiation axis (red solid lines) as well as the components of $\vLhat_N$ (green dashed lines), and compare them to the NR maximum-radiation axis (blue solid lines). 
The construction of the maximum-radiation frame provides Euler angles all the way to the end of the ringdown, that is beyond the end of the EOBNR orbital evolution.   
We have compared the maximum-radiation axis $\boldsymbol{\hat{e}}_{(3)}^R$ as computed in EOBNR and NR for all 70 configurations 
listed in Fig.~\ref{F:SXS_prec_runs}, and found very good agreement, to the level illustrated by case SXS:BBH:0137 in Fig.~\ref{F:axis}. This confirms that the model accurately captures the leading precessional effect that pertains the motion of the orbital plane.

\subsection{Waveforms in the maximum-radiation frame}

 \begin{figure*}
  \includegraphics[angle=0,width=\textwidth, keepaspectratio=true]{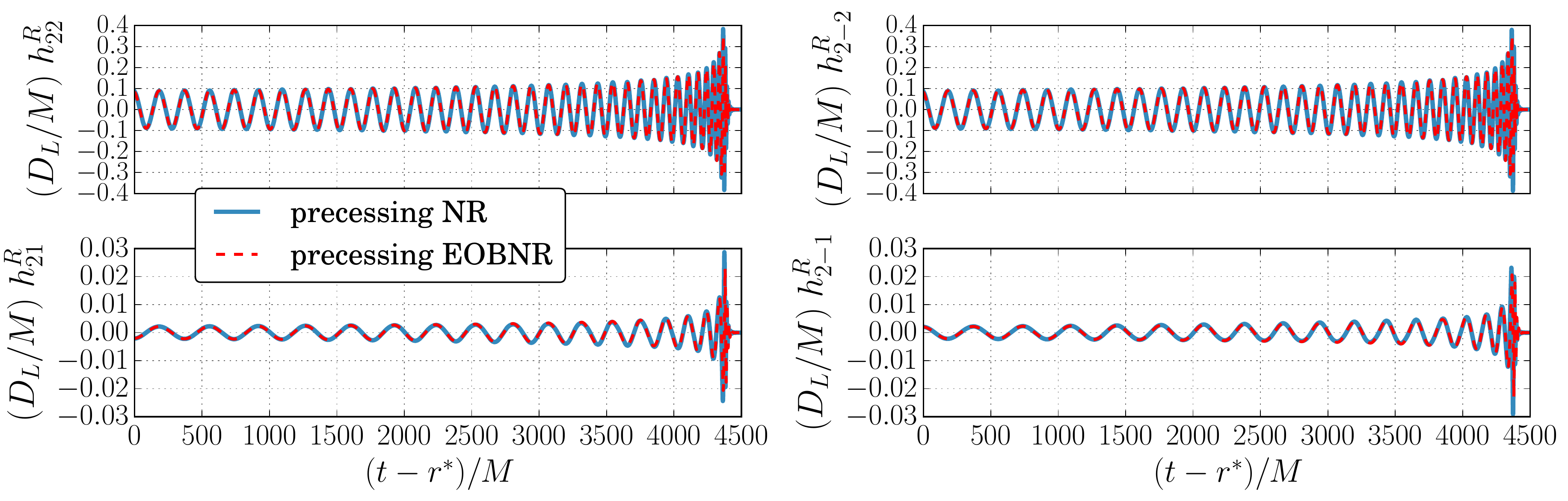}
 \caption{\textbf{Comparison of NR and EOBNR maximum-radiation-frame modes} $h_{2m}^R$ for the same run of Fig.~\ref{F:axis} (SXS:BBH:0137). The NR (EOBNR) curves are shown in solid blue (dashed red). }
 \label{F:Pmodes}
\end{figure*}
Let $h_{\ell m}^R$ be the mode decomposition of $h_+ - i h_{\times}$ in the maximum-radiation frame $\{\boldsymbol{\hat{e}}^R_{(i)}\}$. In the precessing EOBNR model, because of the identity of precessing frame and maximum-radiation frame up to the onset of the ringdown, we have $h_{2 m}^R(t\leq t_{\textrm{match}})=h_{2 m}^P(t\leq t_{\textrm{match}})$. One important assumption in the model is the symmetry $h^P_{2-m}=h_{2m}^{P*}$, which is valid only for nonprecessing binaries. We test this assumption by rotating NR waveforms into the NR maximum-radiation frame. In Fig.~\ref{F:Pmodes} we compare NR and EOBNR maximum-radiation-frame modes for the SXS:BBH:0137 configuration. The inspiral-plunge portion of the signal displays excellent agreement for both positive and negative $m$-modes. 
This level of agreement is common to all 70 waveforms, with overlaps above 99\% for the inspiral part of $m=\pm 2$ modes.  The merger-ringdown portion does not show the same level of accuracy. We remind that the ringdown signal was attached in the attachment frame $\{\boldsymbol{\hat{e}}_{(i)}^J\}$, where modes with different $m$ are mixed by the rotation from the precessing frame. In addition, we  observe that the NR waveforms display some asymmetry between positive and negative $m$-modes; this asymmetry is very mild during the inspiral, but becomes increasingly larger as we approach the merger. The amplitudes of the NR $h^R_{2\pm2}$ modes oscillate around the amplitude of the (symmetric) EOBNR $h^R_{2\pm2}$ modes, and peak at different times. 
The phase difference between the NR $h^R_{2\pm2}$ modes has a secular drift which is typically $\sim 0.2$~rad at merger, with some 
configurations reaching  $\sim 0.6$~rad. The observed disparity between positive and negative modes in NR simulations is discussed in detail in Ref.~\cite{Boyle:2014ioa}.

In order to assess how the small disagreement between the maximum-radiation axes in NR ($\boldsymbol{\hat{e}}^{R,\,{\textrm{NR}}}_{(3)}$) and EOBNR ($\boldsymbol{\hat{e}}^{R,\,{\textrm{EOB}}}_{(3)}$) impacts the waveforms (see Fig.~\ref{F:axis}), we perform the 
following test. We produce two sets of EOBNR inertial-frame modes: (i) one is simply $h_{2m}$, obtained by rotation of the $h_{2m}^P$'s according to the Euler angles in Eq.~(\ref{EulerAngles}), (ii) the other is obtained by rotation of the $h_{2m}^P$'s according to the Euler angles that parametrize the rotation from the NR maximum-radiation frame to the inertial frame of an observer. Then we compare these two sets of EOBNR modes to NR modes in the inertial frame. We find that the $(2,\pm2)$ modes are hardly affected by the discrepancies between $\boldsymbol{\hat{e}}^{R,\,{\textrm{NR}}}_{(3)}$ and $\boldsymbol{\hat{e}}^{R,\,{\textrm{EOB}}}_{(3)}$, while the effect on the $(2,\pm 1)$ modes is slightly larger, especially during the merger. 
We quantify the agreement by computing the sky- and polarization-averaged unfaithfulness with NR waveforms using the modified (NR-based) 
rotation, and find no improvement with respect to the results shown in Fig.~\ref{F:UF} -- several cases are actually worse at high total masses. The overlap  at low total masses is dominated by the inspiral, where we have remarkable agreement between the maximum-radiation-frame modes and between the Euler angles, while at high total masses the main contribution to the unfaithfulness comes from the merger-ringdown signal. Since the EOBNR merger-ringdown is generated in the frame of the remnant spin, the corresponding maximum-radiation axis could be quite different from the NR one.

Finally, we assess the influence of the $(2,0)$ mode in the maximum-radiation frame. While in the EOBNR model, by construction, $h_{20}^R=h_{20}^P=0$, in NR waveforms this mode is nonzero, although at least an order of magnitude smaller than other $\ell=2$ modes. We set the NR $h_{20}^R$ to zero, rotate the NR modes to the inertial frame, and compare them to the original NR inertial-frame modes. In all considered cases, we observe a negligible difference between them
 and no effect on the unfaithfulness, and conclude that it is safe to neglect $h_{20}^P$ in the EOBNR model, at least for the BBH configurations considered in this paper.

\section{Conclusions}
\label{S:concl}
The precessing EOBNR model discussed in this paper was one of the waveform models used in the parameter-estimation study of the first GW observation by LIGO, GW150914~\cite{Abbott:2016izl}. Currently, it is the only waveform model that includes all 15 parameters that characterize a BBH coalescence. In this paper, for the first time, we extensively tested the precessing EOBNR model against 70 NR simulations that span mass ratios from 1 to 5, dimensionless spin magnitudes up to 0.5, and generic spin orientations. While we did not recalibrate the inspiral-plunge signal of the underlying nonprecessing model, we improved the description of the merger-ringdown waveform. In particular, we included different QNMs according to the prograde/retrograde character of the plunge orbit and we prescribed the time of onset of the ringdown according to a robust algorithm that minimizes unwanted features in the amplitude of the waveforms around merger. We introduced a sky- and polarization-averaged unfaithfulness to meaningfully compare precessing waveforms.  We devised a procedure to identify appropriate initial physical parameters for the model given a precessing NR simulation. We found that for Advanced LIGO the precessing EOBNR model has unfaithfulness within about 3\% against the large majority of the 70 NR runs  
when the total mass of the binary varies between $10\msun$ and $200\msun$ and inclinations $\iota = 0, \pi/3, \pi/2$. This means that the model is suitable for detection purposes of these systems. We investigated the GW modes in the maximum-raditation frame, and found very good agreement between NR and precessing EOBNR model 
during the inspiral-plunge part of the waveform. While the merger-ringdown signal is in good agreement with NR in the majority of cases, 
there is still room for future improvements, especially for the $(2,\pm 1)$ modes. 

 No strong statements can be formulated about the size of systematic errors when using precessing EOBNR in the context of parameter estimation. For a NR simulation with parameters within the 90\% credible intervals of GW150914, Ref.~\cite{Abbott:2016izl} shows that precessing EOBNR gives an unbiased measurement of the intrinsic parameters.
 
The precessing EOBNR model is currently being used to infer the astrophysical properties of LVT151012 and GW151226~\cite{TheLIGOScientific:2016pea,Abbott:2016nmj}.

The model presented in this paper is available in the LIGO Algorithm Library (LAL) under the name of \verb+SEOBNRv3+.

\section*{Acknowledgements}
We would like to thank Alejandro Boh\'{e}, Prayush Kumar, Serguei Ossokine, Harald Pfeiffer, and Michael P\"{u}rrer for useful discussions.

\appendix
\section{Rotation formulas for gravitational-wave modes}
\label{S:rot}

In this Appendix we provide basic formulas to transform GW modes from one frame to another. 

Let us consider two orthonormal triads $A$ and $B$ (or reference frames) $\{\vE^A_{(i)}\}$ and $\{\vE^B_{(i)}\}$, with $i=1,2,3$. Let us define the transformation matrix $R_{ij} \equiv \vE^{A}_{(i)} \cdot \vE^{B}_{(j)}$, with $i,j=1,2,3$. $R$ is a rotation matrix that can be parametrized in terms of 3 Euler angles $(\alpha,\beta,\gamma)$. We adopt the \textit{active} point of view: 3-vectors are rotated while the reference frame is unchanged. Let $\{\vE^{A}_{(i)}\}$ be the fixed reference frame, then $(\vE^{A}_{(i)})_j = \delta_{ij}$ and $(\vE^{B}_{(i)})_j = R_{jk}(\vE^{A}_{(i)})_k = R_{ji} = (R^{-1})_{ij}$ and $\vE^B_{(i)}=\vE^A_{(j)}R_{ji}$. Let us represent counterclockwise active rotations by an angle $\varphi$ about the coordinate axes with 
\begin{eqnarray}
R_2(\varphi) &\equiv \left(\begin{array}{ccc}\cos\varphi&0&\sin\varphi\\ 0&1&0\\-\sin\varphi&0&\cos\varphi\end{array}\right)\,,\\
R_3(\varphi) &\equiv \left(\begin{array}{ccc}\cos\varphi&-\sin\varphi&0\\ \sin\varphi&\cos\varphi&0\\0&0&1\end{array}\right)\,.
\end{eqnarray}
Any active rotation (including $R$) can be expressed as $R_3(\alpha)R_2(\beta)R_3(\gamma)$, according to the ZYZ convention for Euler angles:
\begin{widetext}
\be
R(\alpha, \beta, \gamma)=\left(
\begin{array}{ccc}
 \cos \alpha  \cos \beta  \cos \gamma -\sin \alpha 
   \sin \gamma  & -\cos \gamma  \sin \alpha -\cos
   \alpha  \cos \beta  \sin \gamma  & \cos \alpha 
   \sin \beta  \\
 \cos \beta  \cos \gamma  \sin \alpha +\cos \alpha 
   \sin \gamma  & \cos \alpha  \cos \gamma -\cos \beta
    \sin \alpha  \sin \gamma  & \sin \alpha  \sin
   \beta  \\
 -\cos \gamma  \sin\beta& \sin \beta  \sin \gamma 
   & \cos \beta  \\
\end{array}
\right)\,.
\label{E:RotMat}
\en
\end{widetext}
Therefore, if $\beta \ne 0, \pi$,
\be
\label{EulerAngles}
\begin{array}{ccc}
\tan{\alpha} = \cfrac{R_{23}}{R_{13}}\,,&\cos{\beta} = R_{33}\,,& \tan{\gamma} = -\cfrac{R_{32}}{R_{31}}\,.
\end{array}
\en
When $\beta=0$ we have 
\be
\label{EulerAnglesSpecial1}
\tan (\alpha + \gamma) = \frac{R_{21}}{R_{11}}\,,
\en
while when $\beta=\pi$ we have 
\be
\label{EulerAnglesSpecial2}
\tan (\alpha - \gamma) = \frac{R_{21}}{R_{11}}\,.
\en

Let $\boldsymbol{\hat{N}}^A$ be along the line of sight pointing towards the observer; introducing spherical coordinates $\{\boldsymbol{\hat{r}}^A, \boldsymbol{\hat{\theta}}^A, \boldsymbol{\hat{\phi}}^A\}$ in the $A$ frame, we can write $\boldsymbol{\hat{N}}^A \equiv (\sin{\theta^A}\cos{\phi^A},\sin{\theta^A}\sin{\phi^A},\cos{\theta^A})$. So we can define the GW modes in the $A$ frame via
\begin{widetext}
\begin{align}
\label{E:modes}
h_+(t, \theta^A,\phi^A) - ih_{\times}(t,\theta^A,\phi^A) \equiv \sum_{\ell=0}^\infty\sum_{ m=-\ell}^{\ell} h^A_{\ell m}(t)\, _{-2}Y_{\ell m}(\theta^A,\phi^A)\,,\\
h_{\ell m}^A = \int_{-1}^1{\textrm d}(\cos\theta^A)\int_0^{2\pi}{\textrm d}\phi^A \left[h_+(t, \theta^A,\phi^A) - ih_{\times}(t,\theta^A,\phi^A)\right]_{-2}Y^*_{\ell m}(\theta^A,\phi^A)\,,
\end{align}
\end{widetext}
where $_{-2}Y_{\ell m}(\theta^A,\phi^A)$ are $-2$-spin-weighted spherical harmonics. In this paper we only consider modes with $\ell = 2$. Let $\boldsymbol{\hat{N}}^B=R\boldsymbol{\hat{N}}^A \equiv (\sin{\theta^B}\cos{\phi^B},\sin{\theta^B}\sin{\phi^B},\cos{\theta^B})$. The $_{-s}Y_{\ell m}$'s transform according to~\cite{Gualtieri:2008ux,Boyle:2011gg}
\be
_{-s}Y_{\ell m}(\theta^B,\phi^B) = \sum_{m' = -\ell}^\ell {_{-2}Y}_{\ell m'}(\theta^A,\phi^A) D_{mm'}^{(\ell)*}e^{is\zeta}\,,
\label{E:sph_harm_trans}
\en
where $D^\ell_{m m'}$ is the Wigner matrix~\cite{Sakurai:1167961}
\begin{align}
D^{(\ell)}_{m'm}&= e^{-im'\alpha} e^{-im\gamma} \sum_k (-1)^{k-m+m'}\nonumber\\
&\times\cfrac{\sqrt{(j+m)!(j-m)!(j+m')!(j-m')!}}{k!(j+m-k)!(j-k-m')!(k-m+m')!} \nonumber\\
&\times\left[\cos\left(\frac{\beta}{2}\right)\right]^{2j-2k+m-m'} \left[\sin\left(\frac{\beta}{2}\right)\right]^{2k-m+m'}\,,
\label{Wigner}
\end{align}
where $k$ takes values for which the factorials are non-negative, and the factor $e^{is\zeta}$ takes into account the tensorial nature of $_{-s}Y_{\ell m}$, with
\bea
\sin\zeta &= \frac{\sin{\beta}\sin(\phi^A+\gamma)}{\sqrt{ 
1 - [\cos{\beta}\cos{\theta^A} - \cos(\gamma+\beta) \sin{\beta}\sin{\theta^A}]^2}}\,,\\
\cos\zeta &= \frac{\sin{\theta^A} [\cos{\beta} + \cos(\phi^A+\gamma)\cot{\theta^A}\sin{\beta}]}{\sqrt{ 
1 - [\cos{\beta}\cos{\theta^A} - \cos(\gamma+\beta) \sin{\beta}\sin{\theta^A}]^2}}\,.
\ena
Note that the polarizations transform like $h_{+,\times}^B(t,\theta^B,\phi^B) = h_{+,\times}^A(t,\theta^A,\phi^A)e^{-2i\zeta}$, and combining it with Eq.~
 (\ref{E:sph_harm_trans}), we obtain the transformation rule for the modes:
\be
\label{hlmRotation}
h^B_{\ell m} = \sum_{m'=-\ell}^\ell h^A_{\ell m'} D^{(\ell)*}_{m'm} (R^{-1})\,,
\en
where $R^{-1} =  R(-\gamma, -\beta, -\alpha)$ denotes  the inverse rotation.
This expression together with the rotation matrix  (\ref{E:RotMat}) and the Euler angles defined in Eqs.~(\ref{EulerAngles})--(\ref{EulerAnglesSpecial2})
 give complete description of the transformation between two different frames and corresponding GW modes.

\section{Sky- and polarization-averaged overlap}
\label{AppendixA}

Here we define the sky- and polarization-averaged overlap that we employ in Sec.~\ref{S:comparison} to compare precessing NR and EOBNR waveforms. 
We closely follow  Ref.~\cite{Damour:1997ub}, where \textit{min-max} and \textit{max-max} overlaps were first introduced. We review the derivation of those quantities using our notation.

Let us assume that we have a GW signal $s$ and a template $h$ as given in Eqs.~(\ref{E:sstrain})-(\ref{E:hstrain}).
In practice, we do not know the source parameters $(\psi_s, \theta_s, \phi_s)$, and we want to vary the template parameters 
$(\psi_h, \theta_h, \phi_h)$ to maximize the overlap
\be
\mathcal{O} \equiv \frac{(h | s)}{\sqrt{(h|h) (s|s)}} \equiv (\hat{h} | \hat{s} )\,,
\en
with $(\hat{h}|\hat{h})=(\hat{s}|\hat{s})= 1$. The inner product is defined as 
\be
(a | b )  \equiv 4 \Re \int_{0}^{\infty} \frac{\tilde{a}(f)\tilde{b}^*(f)}{S(f)}\,{\textrm d}f\,,
\en
where the tilde denotes the Fourier transform and $S(f)$ is the single-sided noise power spectral density of the detector.
We assume that both signal and template are tapered before taking the Fourier transform to avoid spectral leakage. 
First, we decompose the template into an orthonormal basis
\bea\label{ortho}
\hat{e}_1^h &\equiv& \hat{h}_{+}\,,\\
\hat{e}_2^h &\equiv& \frac{ \hat{h}_{\times} -  ( \hat{h}_{\times} |  \hat{h}_{+})  \hat{h}_{+}}
{\sqrt{ 1 -  ( \hat{h}_{+} |  \hat{h}_{\times})^2}}\,.
\ena
Using this basis we can write the template as 
\be
\hat{h} = \hat{e}^h_1\cos\alpha  + \hat{e}^h_2\sin\alpha \,,
\en 
where
\bea
\cos\alpha &\equiv& \frac{F_{+} |h_{+}| - F_{\times}|h_{\times}| (\hat{h}_{+} | \hat{h}_{\times})}{\sqrt{(h|h)}}\,, \\
\sin\alpha &\equiv& \frac{F_{\times}|h_{\times}| \sqrt{1 - (\hat{h}_{+} | \hat{h}_{\times})^2}}{\sqrt{(h|h)}}\label{alpha}\,.
\ena
Similarly (by replacing $h \curvearrowright s$ and $\alpha \curvearrowright \beta$ in Eqs.~(\ref{ortho})-(\ref{alpha})), we can write the signal as
\be
\hat{s} = \hat{e}_1^s\cos\beta  + \hat{e}_2^s\sin\beta\,.
\en
The overlap is now a function of $\alpha$ and $\beta$. In turn, $\alpha$ and $\beta$ are functions of the polarization angles and the sky location. Let us emphasize that $\beta(\psi_s, \theta_s, \phi_s)$  is unknown. Maximizing the overlap over the template parameters we have
\be
\max_{\alpha}{\mathcal{O}} = \sqrt{ (\hat{s} | \hat{e}_1^h)^2 + (\hat{s} | \hat{e}_2^h)^2 }\,.
\label{E:max_a}
\en
The square of $\max_{\alpha}\mathcal{O}$ describes an ellipse parametrized by the angle $\beta$.  The quantities 
\textit{min-max} and \textit{max-max} introduced
in Ref.~\cite{Damour:1997ub} correspond to the minimum (semi-minor axis) and maximum (semi-major axis) of Eq.~(\ref{E:max_a})
over all possible values of $\beta$:
\begin{align}
\min_{\beta} \max_{\alpha}{\mathcal{O}} &= \frac{1}{\sqrt{2}}\sqrt{A+B - \sqrt{(A-B)^2 + 4C^2}}\,,\\
\max_{\beta} \max_{\alpha}{\mathcal{O}} &= \frac{1}{\sqrt{2}}\sqrt{A+B + \sqrt{(A-B)^2 + 4C^2}}\,,
\end{align}
where
\begin{align}
A &\equiv (\hat{e}_1^h |  \hat{e}_1^s)^2 + (\hat{e}_2^h |  \hat{e}_1^s)^2\,, \\
B &\equiv (\hat{e}_1^h |  \hat{e}_2^s)^2 + (\hat{e}_2^h |  \hat{e}_2^s)^2\,, \\
C &\equiv (\hat{e}_1^h |  \hat{e}_1^s)(\hat{e}_1^h |  \hat{e}_2^s) + (\hat{e}_2^h |  \hat{e}_1^s) (\hat{e}_2^h |  \hat{e}_2^s)\,.
\end{align}
Of course, for a generic polarization and sky location, the overlap (\ref{E:max_a}) lies between the values of min-max and max-max. However, the range between min-max and max-max is often quite large, and we do not know the distribution of overlaps therein as we vary the polarization angle and the sky location of the source. Thus, we introduce a new quantity, the \textit{sky- and polarization-averaged overlap}, or \textit{average-max}: this is the overlap (\ref{E:max_a}) averaged over all possible polarization angles and sky locations, so it is the mean of the distribution of overlaps bounded by min-max and max-max
\begin{widetext}
\bea
{\textrm{average-max}} \equiv
\frac{1}{8 \pi^2} \int_0^{2\pi} {\textrm d}\psi_s  \int_{-1}^1 {\textrm d}(\cos\theta_s)\int_0^{2\pi} {\textrm d}\phi_s
\sqrt{ (\hat{s} | \hat{e}_1^h)^2 + (\hat{s} | \hat{e}_2^h)^2 }\,. 
\label{E:averOlap}
\ena
\end{widetext}

\section{Ringdown spectrum}
\label{AppendixB}

In this Appendix we describe how we improve the EOB ringdown spectrum with respect to Ref.~\cite{Pan:2013rra}. 

Let us consider an isolated Kerr BH of mass $M_f$ and spin $\boldsymbol{\chi}_f$.
We denote the complex QNM frequency with $\sigma_{\ell m n}$, where $(\ell,m)$ are spheroidal-harmonic indices and $n$ is the overtone index ($n = 0,1, \cdots$). We set $\sigma_{\ell m n} \equiv \omega_{\ell m n} - i/\tau_{\ell m n}$, where $\omega_{\ell m n} > 0$ is the real QNM frequency and $\tau_{\ell m n} > 0$ is its decay time. Note that $M_f\sigma_{\ell m n}$ depends on\footnote{Here, we assume that $\boldsymbol{\chi}_f$ is aligned or antialigned with the $z$-axis, which is used to perform the spheroidal harmonic decomposition.} $\pm \chi_f $ and $M_f\sigma_{\ell -m n} (\chi_f) = M_f\sigma_{\ell m n} (-\chi_f)$. 

In the precessing EOBNR model we prescribe that the QNM content of the spectrum varies according to the character of the EOBNR trajectory at merger (i.e., at time $t_{\textrm{match}}$). We consider an EOBNR orbit to be prograde (retrograde) if $\boldsymbol{\hat{\chi}}_f\cdot \vLhat_{\textrm{match}} > 0$ ($\boldsymbol{\hat{\chi}}_f\cdot\vLhat_{\textrm{match}} < 0$), where $\vLhat_{\textrm{match}}\equiv \vLhat(t_{\textrm{match}})$. The time $t_{\textrm{match}}$ is defined in defined in Appendix~\ref{AppendixC}.

For prograde EOBNR orbits we set
\begin{align}
M_f \sigma_{2mn} = \left\{
\begin{array}{ll}
      M_f\sigma_{2mn}(\chi_f)\,, & {\rm if }\;m > 0\,, \\
      -M_f\sigma^*_{2-mn}(\chi_f)\,, & {\rm if }\;m < 0\,.
         \end{array}
    \right.
\end{align}
When $m=0$, we use 
\begin{align}
M_f \sigma_{20n} = \left\{
\begin{array}{ll}
      M_f\sigma_{20n}(\chi_f)\,, & {\rm if }\; n =0,\cdots,4\,, \\
      -M_f\sigma^*_{20(n-5)}(\chi_f)\,, & {\rm otherwise }\,.
         \end{array}
    \right.
\end{align}

For retrograde EOBNR orbits we set
\begin{align}
M_f \sigma_{2mn} = \left\{
\begin{array}{ll}
      -M_f\sigma^*_{2mn}(-\chi_f)\,, & {\rm if }\; m > 0\,, \\
      M_f\sigma_{2-mn}(-\chi_f)\,, & {\rm if }\; m < 0\,,
         \end{array}
    \right.
\end{align}
When $m=0$, we use 
\begin{align}
M_f \sigma_{20n} = \left\{
\begin{array}{ll}
     - M_f\sigma^*_{20n}(-\chi_f)\,, & {\rm if }\; n =0,\cdots,4\,, \\
      M_f\sigma_{20(n-5)}(-\chi_f)\,, & {\rm otherwise }\,.
         \end{array}
    \right.
\end{align}

In order to improve the stability of the ringdown attachment, for all EOBNR orbits that are precessing (i.e., with the exception of spin-aligned BBHs) and for very asymmetric binaries ($\nu < 0.05$) we introduce $\sigma_{2\mp20}$ in the ringdown spectrum of $(2,\pm2)$ modes. For retrograde EOBNR orbits and very asymmetric binaries ($\nu < 0.1$) we introduce $\sigma_{2\mp10}$ in the ringdown spectrum of $(2,\pm1)$ modes.
These prescriptions are also motivated by what is observed in merger-ringdown Teukolsky-code--based waveforms in the test-particle limit (QNM mixing)~\cite{Barausse:2011kb,Taracchini:2014zpa}.

Similarly to what is done in the underlying nonprecessing EOBNR model~\cite{Taracchini:2013rva}, pseudo-QNMs are introduced to bridge the gap between the end-of-inspiral and the least-damped-QNM frequency, especially for large mass ratios and large spin magnitudes.

\section{Time of onset of the ringdown}
\label{AppendixC}
In this Appendix we explain how the time of onset of the ringdown signal (i.e., $t_{\textrm{match}}$) is chosen in the precessing EOBNR model.

In the nonprecessing model, $t_{\textrm{match}} = t_{\textrm{peak}}^{22}$, where $t_{\textrm{peak}}^{22}$ is the time when the $(2,2)$-mode amplitude peaks; by virtue of the iterative computation of the nonquasicircular corrections (see Eqs.~(21)-(25) of Ref.~\cite{Taracchini:2012ig}), $t_{\textrm{peak}}^{22}=t_{\textrm{peak}}^{\Omega} + \Delta t^{22}_{\textrm{peak}}$, where $t_{\textrm{peak}}^{\Omega}$ is the time when the orbital frequency peaks (which typically occurs close to the light-ring crossing) and $\Delta t^{22}_{\textrm{peak}}$ is a prescribed function of the BBH parameters motivated by Teukolsky-code--based studies in the test-particle limit~\cite{Barausse:2011kb,Taracchini:2014zpa}. For a precessing binary, we apply the nonquasicircular corrections that a nonprecessing binary with the same mass ratio $q$ and initial spins $\boldsymbol{\chi}_i(0) \cdot \vLhat_N(0)$ would have. The corrections are applied only to the precessing-frame $h^P_{2\pm 2}$ modes.

We attach the ringdown part of the signal in the frame associated with the final angular momentum of the system, where the hierarchy among the 
modes is not strongly present. In general, the attachment-frame modes $h_{2 m}^J$ reach globally maximum amplitude at a time $t^{2m, J}_{\textrm{peak}}$ that is different from $t_{\textrm{peak}}^{\Omega} + \Delta t^{22}_{\textrm{peak}}$. Moreover, several local maxima in $|h^J_{2 m}|$ can appear due to: (i) conservative spin-spin couplings~\cite{Buonanno:2010yk},  (ii) mixing of modes with different $m$ because of the rotation between frames. We also find that some binary configurations do not reach any peak in $\Omega$ or $|h^J_{2 m}|$ before the termination of the orbital dynamics. All this makes the choice of an attachment time more complicated than in the nonprecessing case.  

Therefore, we proceed as follows. First, for nonspinning BBHs the EOBNR light-ring is located very close to the peak of $\Delta_u/r^2$, where $\Delta_u(r)$ is the radial potential (see Eq.~(2) of Ref.~\cite{Taracchini:2013rva}), that corresponds to the ``$00$'' component of the effective metric. Let $t_{\textrm{peak}}^{\Delta_u/r^2}$ be the time when $\Delta_u/r^2$ peaks. We define
\bea
t_{\textrm{max}}^{\Omega} \equiv \left\{ 
   \begin{array}{ll}
     t^{\Omega}_{\textrm{peak}}\,, &{\rm if }\,\, \Omega \,\, {\rm peaks}\,,\\ 
     t_{\textrm{peak}}^{\Delta_u/r^2}\,,&{\rm otherwise}\,.
   \end{array}
   \right.
\ena
Then, we denote with $t^{2m,J}_{\textrm{flat}}$ the time when ${\rm d}|h_{2m}^J|/{\rm d}t$ has a minimum (close to merger), and we define
\bea
t^{2m}_{\textrm{max}} \equiv\left\{ 
   \begin{array}{ll}
   t^{2m,J}_{\textrm{peak}}\,,&{\rm if }\,\, |h_{2m}^J| \,\,{\rm peaks}\,,\\
   t^{2m,J}_{\textrm{flat}}\,, & {\rm otherwise}\,.
   \end{array}
 \right.
 \ena
Finally, we set
\be \label{tmatch}
t_{\textrm{match}} = \min{(t_{\textrm{max}}^{\Omega}+\Delta t_{\textrm{peak}}^{22},t_{\textrm{max}}^{22,J})}\,,
\en
where for simplicity we use the same attachment time for all values of $m$. For a vast part of the parameter space we can employ Eq.~(\ref{tmatch}), and we can achieve a plateau in $|h_{2m}^J|$ around the time of attachment. However, we occasionally encounter sudden variations in the frequency and/or  
amplitude of the attachment-frame inspiral-plunge modes. To improve the robustness of the attachment procedure and 
achieve a plateau in $|h_{2\pm2}^J|$ around the time of attachment, we shift the attachment time $t_{\textrm{match}}$ around the value predicted by Eq.~(\ref{tmatch}) and minimize
\be
\sum_{m=\pm2}{\left(\max_{t\leq t_{\textrm{match}}} |h_{2m}|-\max_{t> t_{\textrm{match}}} |h_{2m}|\right)^2}\,.
\en
This choice is motivated by empirical evidence collected studying the evolution of the inspiral-plunge modes in  
the precessing NR waveforms.
\bibliographystyle{apsrev}
\bibliography{references}
\end{document}